\documentclass[aip,reprint,superscriptaddress,amsmath,amssymb, 
longbibliography,nobibnotes,nofootinbib,altaffillsymbol,twocolumn]{revtex4-2} 
\usepackage{amsmath} 
\usepackage{amssymb} 
\usepackage{amsfonts} 
\usepackage{bm} 
\usepackage{graphicx} 
\usepackage{dcolumn} 
\usepackage{txfonts} 
\usepackage{makeidx}
\usepackage{color}
\usepackage{mathtools}
\usepackage{threeparttable}
\usepackage[linkcolor=blue,anchorcolor=black,citecolor=blue,colorlinks=true]{hyperref}
\usepackage{breakurl}
\usepackage{multirow}
\usepackage{array}
\usepackage{xr}  
\usepackage[T1]{fontenc}    
\usepackage{caption}
\usepackage{upgreek}   

\usepackage{xpatch}
\makeatletter
\xpatchcmd{\@ssect@ltx}{\@xsect}{\protected@edef\@currentlabelname{#8}\@xsect}{}{}
\xpatchcmd{\@sect@ltx}{\@xsect}{\protected@edef\@currentlabelname{#8}\@xsect}{}{}
\makeatother

\newcommand{\degC}{{\rm ^\circ C}}

\newcommand{\gXe}{\gamma_{\rm Xe}}
\newcommand{\GXe}[1]{\Gamma_{\rm #1}}
\newcommand{\bfr}{\mathbf{r}}
\newcommand{\rme}{{\rm e}}
\newcommand{\rmd}{{\rm d}}
\newcommand{\rmi}{{\rm i}}
\newcommand{\XeName}[1]{$\rm ^{#1}Xe$}

\begin{document}
\title{Frequency Shift Caused by Nonuniform Field and Boundary Relaxation in Magnetic Resonance and Comagnetometers}
\author{Xiangdong Zhang}
\affiliation{Beijing Computational Science Research Center, Beijing 100193, People's Republic of China}%
\affiliation{College of Physics and Optoelectronic Engineering, Shenzhen University, Shenzhen 518060, China}%
\author{Jinbo Hu}
\affiliation{Beijing Computational Science Research Center, Beijing 100193, People's Republic of China}%
\author{Da-Wu Xiao}
\affiliation{Beijing Computational Science Research Center, Beijing 100193, People's Republic of China}%
\author{Nan Zhao}
\email{nzhao@csrc.ac.cn}
\affiliation{Beijing Computational Science Research Center, Beijing 100193, People's Republic of China}%
\date{\today}

\begin{abstract}
  In magnetic resonance experiments, 
  it is widely recognized that a nonuniform magnetic field can lead to an increase in the resonance line width, as well as a reduction in sensitivity and spectral resolution. However, a nonuniform magnetic field can also cause shifts in resonance frequency, which has received far less attention. 
  In this work, we investigate the frequency shift caused by boundary relaxation and nonuniform magnetic field with arbitrary spatial distribution.
  We find that this frequency shift is spin-species dependent, implying a systematic error in NMR gyroscopes and comagnetometers. 
  The first order correction to this systematic error is proportional to the difference of boundary relaxation rate, and dominates for small cells. 
  In contrast, the third and higher order corrections arise from the difference of gyromagnetic ratios of spin species, and dominates for large cells. 
  This insight helps understanding the unexplained isotope shifts in recent NMR gyroscopes and new physics searching experiments that utilize comagnetometers. 
  Finally, we propose a tool for wall interaction research based on the frequency shift's dependency on boundary relaxation. 
\end{abstract}

\maketitle

\section*{Introduction}
The nuclear magnetic resonance (NMR) technique has important applications in many fields, such as biochemistry~\cite{James1975, Mlynarik2017}, medical imaging~\cite{Henderson1983, Hendee1999}, inertial navigation~\cite{Meyer2014, Walker2016}, and new physics detection using comagnetometers~\cite{Yevgeny2017, Abel2017, Safronova2018, Graham2018, Wu2019, Terrano2022, JacksonKimball2023}. 
For gas-phase NMR, the interaction between diffusion effect and nonuniform magnetic field has been a focus of research for decades. A nonuniform magnetic field contributes an extra relaxation rate to the diffusive nuclear spins, which is usually harmful to the precision of an NMR experiment. Extensive works have been done to analyze this extra relaxation rate in various scenarios~\cite{Schearer1965, Cates1988, Cates1988a, McGregor1990, Stoller1991, Zielinski2000, Zhao2008a, Golub2011, Zheng2011, Zhan2020, Lee2021b}. 
On the contrary, the frequency shift caused by a nonuniform magnetic field is relatively underexplored. 
An unexpected spin-species dependent frequency shift caused by a nonuniform magnetic field can introduce significant systematic error in NMR based precision measurements that need to suppress the main field fluctuation by comparing the resonance frequencies of different spins. 
Such measurements recently show great potential in the exploration of new physics~\cite{Bulatowicz2013, Tullney2013, Zimmer2018, Lee2018, Sachdeva2019, Wu2019, Terrano2022, Feng2022, Zhang2023a, Bloch2023}. 
So, it is of great interest to understand how a nonuniform magnetic field affect the resonance frequency of spins. 


Nonuniform magnetic fields of different magnitude can result in significantly different behaviors. 
With a small nonuniform field, the structure of the free-induction decay (FID) spectrum is not changed, and perturbation theory can be used to estimate the frequency shift. This region is called the fast-diffusion area, meaning that the spins can diffuse throughout the whole cell before completely decayed. 
When the field inhomogeneity is large compared to the diffusion speed, the FID spectrum can split due to the symmetry breaking of eigenmodes~\cite{Stoller1991,Zhang2023}. This splitting phenomenon lies in the intermediate regime between fast- and slow-diffusion limit, and is closely related to the well known edge enhancement effect~\cite{Deswiet1995, Saam1996, Song1998, Zhao2010}. 
This work focuses on the small nonuniform field region, as most of the NMR based precision measurements lie in this region. 

In an early work of Cates {\it et al.}~\cite{Cates1988}, the frequency shift $\delta \omega$ of diffusive spins filled in a spherical container of radius $R$ is derived by applying the second order perturbation theory on the Torrey equation:
\begin{equation}
  \delta \omega \approx \frac{\gamma R^2}{10 B_0} \left( 
     \left|  \nabla B_x \right|^2 + \left|  \nabla B_y \right|^2
  \right),
  \label{Eq:freq_shift_cates1988}
\end{equation}
where $\gamma$ is the gyromagnetic ratio of spins, $B_0$ is the uniform main field along $z$ direction and $B_x, B_y$ are the transverse fields. 
This formula is valid for fast-diffusion and high-pressure limit. 
Similar results were also presented in later works based on the Redfield theory~\cite{Zheng2011, Guigue2014, Jeener2015} and in this work (Eq.~\eqref{Eq:freqShift_linG_1st}). 
Unfortunately, this formula requires the magnetic field to hold an odd-parity symmetry (see \ref{sec:adx_odd_parity} for the meaning of odd-parity symmetry). If the odd-parity symmetry is broken, as in many practical experiments, this formula will underestimate the magnitude of the frequency shift. 

On the other hand, Sheng {\it et al.}~\cite{Sheng2014} analyzed the frequency shift due to an arbitrary nonuniform field by numerical simulation. They expanded the nonuniform magnetic field using spherical harmonics and transformed the Torrey equation into a linear ODE system, which is suitable for numerical simulation. Through dimensional analysis, the frequency shift caused by a quadratic gradient field was found to be 
\begin{equation}
  \delta \omega \propto \frac{\gamma^3 G_2^3 R^{10}}{D^2}, 
  \label{Eq:freq_shift_sheng2014}
\end{equation}
where $D$ is the diffusion constant and $G_2$ is the strength of quadratic gradient. 
Due to the dependence on $\gamma$ and $D$, this frequency shift will lead to a systematic error in $\rm ^{3}He$-\XeName{129} comagnetometer experiments~\cite{Sheng2014}. We will show later that Eq.~\eqref{Eq:freq_shift_sheng2014} is actually the third order perturbation correction of frequency shift for quadratic gradient field (Eq.~\eqref{Eq:freqShift_quadG_3rd}).  

Isotope shift, a major systematic error in comagnetometer and NMR gyroscope setups that use \XeName{129} and \XeName{131} spins, is closely relevant to the topic of this work. 
Bulatowicz {\it et al.}~\cite{Bulatowicz2013} first reported this isotope shift phenomenon in an Axion searching experiment, and showed that this systematic error is directly related to the effective magnetic field of polarized alkali atom spins (alkali field)~\cite{Schaefer1989} generated by the collision between xenon and alkali-metal atoms. 
They suppose that this isotope shift is originated from the possible difference of collisional enhancement factor of \XeName{129} and \XeName{131}. 
Later works~\cite{Vershovskii2018,Petrov2020,Vershovskii2020,Petrov2022} attribute this isotope shift to the incomplete averaging of alkali field by the two xenon isotopes, which have different relaxation time $T_2$. 
With reference to the above works, this work will start from a new perspective -- the diffusion equation, and try to quantitatively elucidate the origin of isotope shift. 

\begin{figure*}[ht]
  \includegraphics[width=\textwidth]{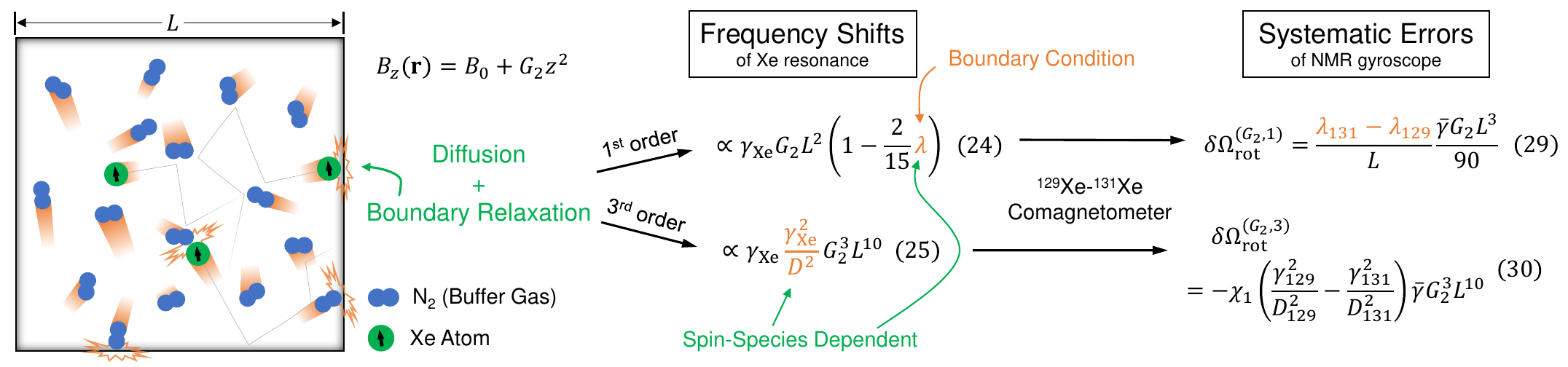}
  \caption[]{\label{Fig:cartoon_demo}{\textbf{Cartoon of the frequency shift mechanism.} 
  Considering the diffusion motion and boundary relaxation of Xe spins, a nonuniform magnetic field can shift the resonance frequency of Xe spins. 
  The first order frequency correction slightly depends on the boundary condition $\lambda$, while the third order correction is proportional to the ratio of gyromagnetic ratio $\gXe$ and diffusion constant $D$. 
  Both $\lambda$ and $\gXe^2 / D^2$ can be different for different Xe isotopes, thus will introduce systematic error in comagnetometer type experiments. 
  As an example, the systematic errors introduce to NMR gyroscopes, a kind of \XeName{129}-\XeName{131} comagnetometer, are shown by $\delta \Omega_{\rm rot}^{(G_2,1)}$ and $\delta \Omega_{\rm rot}^{(G_2,3)}$. 
  These systematic errors are generally not negligible, limiting the absolute accuracy of comagnetometers as well as the long time stability of NMR gyroscopes. 
  On the other hand, the factor $(\lambda_{131}-\lambda_{129})$ in $\delta \Omega_{\rm rot}^{(G_2,1)}$ enables a new tool for boundary relaxation rate measurement, which should be much faster than previously reported methods. 
  Refer to Eqs.~\eqref{Eq:freqShift_quadG_1st}, \eqref{Eq:freqShift_quadG_3rd}, \eqref{Eq:deltaOmega_quadG_1st} and \eqref{Eq:deltaOmega_quadG_3rd} for more details.}}
\end{figure*}

In this work, we aim to investigate the frequency shift effect of a nonuniform magnetic field with arbitrary spatial distribution, and especially its interaction with boundary relaxation. 
Perturbation corrections up to third order are derived to analytically illustrate the sources of frequency shift and isotope shift. 
Figure~\ref{Fig:cartoon_demo} shows a sketch for the core mechanism that leads to a spin-species dependent frequency shift. 
By saying a frequency shift is spin-species dependent, we mean the equivalent magnetic field of this frequency shift is different for different types of spin. Thus, this frequency shift cannot be cancelled together with the main field fluctuation in comagnetometer type experiments.
For linear gradient magnetic field, whose spatial distribution has odd-parity, we find that the first order correction to the frequency shift (Eq.~\eqref{Eq:freqShift_linG_1st}) is similar to Eq.~\eqref{Eq:freq_shift_cates1988}, which is inversely proportional to the main field $B_0$ and proportional to the square of transverse fields. 
On the other hand, for quadratic gradient field, whose distribution does not have odd-parity, the first order correction to frequency shift (Eq.~\eqref{Eq:freqShift_quadG_1st}) is proportional to the nonuniform part of the longitudinal field and does not have a $B_0$ in the denominator, which is significantly different from Eq.~\eqref{Eq:freq_shift_cates1988}. This result shows that Eq.~\eqref{Eq:freq_shift_cates1988} may underestimate the magnitude of the frequency shift caused by nonuniform magnetic field in experiments, especially in NMR experiments that use a large main field. 
We also find that Eq.~\eqref{Eq:freq_shift_sheng2014} is actually the third order correction of frequency shift for quadratic gradient field (Eq.~\eqref{Eq:freqShift_quadG_3rd}). Due to the $\gamma^3/D^2$ and $R^{10}$ dependence, this third order correction is spin-species dependent and contributes significant isotope shift for large cells. 

The boundary relaxation is another crucial factor that not well studied previously but can influence the resonance frequency by the interaction with nonuniform field. We find that the boundary relaxation will lead to a small correction to the first order frequency shift, and this correction is spin-species dependent due to the wall interaction mechanism difference of different spins. 
This effect of boundary relaxation is vital for comagnetometer experiments where the first order correction of frequency shift can completely cancel if boundary condition is not considered.
We show that the isotope shifts caused by the first and third order corrections of frequency shift (Eqs.~\eqref{Eq:deltaOmega_linG_1st}-\eqref{Eq:deltaOmega_quadG_3rd}) are both important and have distinct behaviors. The first order correction to isotope shift is proportional to the difference of boundary relaxation and is more important for small cells, while the third order correction only weakly depends on boundary relaxation and is more important for large cells. 
The systematic errors introduced by these corrections is fatal in new physics searching experiments where the absolute value of frequency ratio are of great concern.  
As a reference, the systematic error for the rotation signal of an NMR gyroscope can be as large as $10~{\rm \mu Hz}$ in typical experiment conditions (see Fig.~\ref{Fig:compare_systematicErrors}). 
This insight can help explain the isotope shift effect~\cite{Bulatowicz2013,Vershovskii2018,Petrov2020,Vershovskii2020,Petrov2022} in recent NMR gyroscopes and new physics searching experiments, in which alkali field is a highly nonuniform effective-magnetic-field~\cite{Schaefer1989}. 
Furthermore, based on the derived systematic error formulas, a method for fast wall relaxation measurement is proposed.




\section*{Results and Discussion}

In this work, we will focus on the high-pressure and fast-diffusion limit, which is commonly satisfied in recent NMR based precision measurement experiments. 
The high-pressure limit requires a large main field and a small diffusion constant (i.e. $\gamma B_0 L^2 / D \gg 1$~\cite{Zheng2011}), allowing the application of Rotating Wave Approximation (RWA) to the Torrey equation. 
The fast-diffusion limit requires a slow relaxation rate compared to the diffusion speed (i.e. $D T_2 / L^2 \gg 1$~\cite{Zheng2011}), so that the spin evolution can be well approximated by a single eigenmode and the effect of nonuniform magnetic field can be treated with perturbation theory. 
Weak boundary relaxation ($\lambda \ll 1$) is also assumed based on practical experiment conditions.

\subsection*{Torrey Equation}
Use Xe nuclear spin as an example. In experiments utilizing spin-exchange optical pumping technique~\cite{Walker1997}, such as NMR gyroscopes and Rb-Xe comagnetometers, the diffusion of Xe nuclear spins can be described by the Bloch-Torrey equation~\cite{Torrey1956, Appelt1998, Seroussi2017}
\begin{eqnarray}
        \frac{\partial \mathbf{M}(\mathbf{r},t)}{\partial t} &=& D\nabla^2 \mathbf{M}(\mathbf{r},t) 
        -\gamma_{\rm Xe} \mathbf{B}(\mathbf{r})\times\mathbf{M}(\mathbf{r},t)\notag \\
        &&-\mathbf{\Gamma}_0\cdot\mathbf{M}(\mathbf{r},t) + R_{\rm p}' \left[ \mathbf{S}(\mathbf{r}) - \mathbf{M}(\mathbf{r}, t)\right],
    \label{Eq:BlochEquation}
\end{eqnarray}
where $\mathbf{M}(\mathbf{r},t)$ is the Xe nuclear spin magnetization, 
$D$ is the diffusion constant of Xe atoms, 
$\gamma_{\rm Xe}$ is the gyromagnetic ratio of Xe nuclear spins,
$\mathbf{B}(\mathbf{r})$ is the magnetic field distribution,
$\mathbf{\Gamma}_0 = \Gamma_{\rm 20}(\hat{x}\hat{x}+\hat{y}\hat{y})+\Gamma_{\rm 10}\hat{z}\hat{z} $ 
is a tensor describing the transverse and longitudinal spin relaxation processes, 
and the last term arises from the spin-exchange pumping process between Xe and alkali atom spins, with $R_{\rm p}'$ the spin-exchange pumping rate, and $\mathbf{S}(\mathbf{r})$ the alkali electron spin magnetization. 
Here, we assume that the alkali electron spins are polarized along the $\hat{z}$ direction, i.e., $\mathbf{S}(\mathbf{r})=S_z(\mathbf{r}) \hat{z}$.

To further simplify Eq.~\eqref{Eq:BlochEquation}, we assume that the magnetic field only has $\hat{z}$ component (the effect of transverse component can be estimated using Eq.~\eqref{Eq:def_btot}), i.e., 
\begin{eqnarray}
   \mathbf{B}(\mathbf{r})=[B_0 + B_1(\mathbf{r})]\hat{z}, 
   \label{Eq:assumptionOfBz}
\end{eqnarray}
with $|B_1(\mathbf{r})| \ll B_0$. 
The homogeneous field $B_0$ defines the main precession frequency $\Omega_0 = - \gamma_{\rm Xe} B_0$,
while the nonuniform field $B_1(\mathbf{r})$, which may originate from the imperfection of the coils, the environmental stray fields, and the alkali field, can contribute to spin relaxation and frequency shift. 

With assumption Eq.~\eqref{Eq:assumptionOfBz}, the transverse and longitudinal components of Eq.~\eqref{Eq:BlochEquation} are decoupled.
The equation of motion of the transverse components $M_\pm(\mathbf{r},t) \equiv M_x(\mathbf{r},t) \pm \rmi M_y(\mathbf{r},t)$ is  
\begin{equation}
    \frac{\partial M_{\pm }(\mathbf{r}, t)}{\partial t}= \left[D \nabla^2 - \Gamma_{\rm 2c} \mp \rmi \gamma_{\rm Xe} B_z(\mathbf{r}) \right]M_{\pm}(\mathbf{r},t),
    \label{Eq:TransverseEoM}
\end{equation}
where $\Gamma_{\rm 2c} \equiv \Gamma_{20} + R_{\rm p}'$, 
and $B_z(\mathbf{r}) \equiv B_0+B_1(\mathbf{r})$. 
Below, we will focus on Eq.~\eqref{Eq:TransverseEoM}, which determines the FID behavior of Xe spins. 

\subsection*{Boundary Condition}
The boundary condition of Eq.~\eqref{Eq:TransverseEoM} is determined by the spin relaxation processes due to wall interaction. 
A perturbation treatment together with kinetic theory gives the following boundary condition of the Xe nuclear spin density matrix $\rho(\mathbf{r}, t)$~\cite{Wu2021}
\begin{equation}
    \left.\left[\mathbf{n}\cdot \nabla \rho(\mathbf{r}, t) + \hat{\mu} \rho(\mathbf{r}, t)\right]\right\vert_{\mathbf{r}\in \partial V}=0,
\end{equation}
where $\partial V$ is the boundary of solution domain $V$, $\mathbf{n}$ is the normal vector of the boundary (pointing outward),
and $\hat{\mu}$ is an operator reflecting the wall-interaction induced transitions between different components of the Xe spin polarization.

\begin{figure}
  \includegraphics[scale=1]{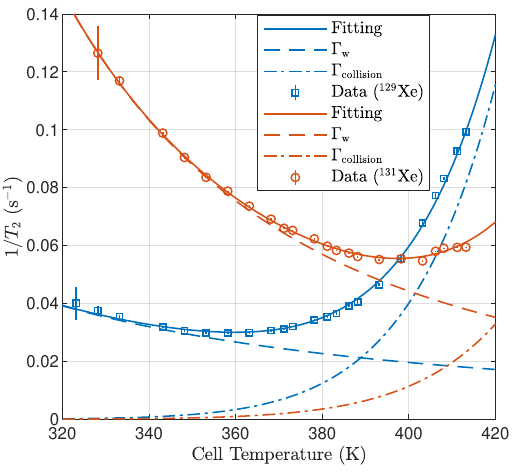}
  \caption[]{\label{Fig:boundaryCond_measurement}{\textbf{A demo of Xe wall relaxation measurement.} 
  The transverse relaxation rate $1/T_2$ of Xe spins mainly consists of the wall relaxation rate $\Gamma_{\rm w}$ and the collisional relaxation rate $\Gamma_{\rm collision}$. Due to the competition between wall trapping and thermal motion, $\Gamma_{\rm w}$ decreases as temperature rises. 
  On the other hand, $\Gamma_{\rm collision}$ is proportional to the density of alkali atoms (e.g. Rb), and thus increases as temperature rises. 
  By measuring the temperature dependence of $1/T_2$, we can extract the wall relaxation rate of Xe spins through nonlinear least squares fitting. 
  This measurement is performed on a RbH coated 8~mm cubic cell (Cell ID: CH7) containing 3.6~Torr \XeName{129}, 35.6~Torr \XeName{131}, 167~Torr $\rm N_2$, 6~Torr $\rm H_2$ and a small droplet of Rb metal. 
  The transverse relaxation rate of \XeName{129} and \XeName{131} spins is measured by fitting the decaying rate of FID signal, using the experiment setup of our previous work~\cite{Zhang2023}. Solid and dashed lines are fitting curves using the model in section "\nameref{sec:adx_wallRelax_measurement}" in the methods. The boundary conditions for \XeName{129} and \XeName{131} at $110\degC$ are estimated to be $\lambda_{129} = \left. (5.3 \pm 2.0) \times 10^{-3} \right.$ and $\lambda_{131} = \left. (13.0 \pm 3.8) \times 10^{-3} \right.$ using these fittings.
  Data points in the figure are given by a Lorentzian fitting as shown in Fig.~S1(g) of our previous work~\cite{Zhang2023}. The error bar represents the 95\% confidence interval returned by the fitting algorithm.
  See Supplementary Data 1 for the source data of this figure.}}
\end{figure}

We replace the operator $\hat{\mu}$ with a constant number for simplification, leading to the following boundary condition:
\begin{equation}
    \left.\left(\mathbf{n}\cdot \nabla M_{\pm}(\mathbf{r},t) + \frac{\lambda}{L}\cdot M_{\pm}(\mathbf{r},t) \right) \right\vert_{\mathbf{r} \in \partial V} = 0,
    \label{Eq:BoundaryConditions}
\end{equation}
where $\lambda \ge 0$ is a dimensionless constant describing the depolarization strength on the container wall and $L$ is a linear size of $V$ (e.g. the side length for cubic $V$ or the radius for spherical $V$, the choice of $L$ does not affect the physics). 
When $\lambda \rightarrow 0$, the boundary condition becomes $\mathbf{n}\cdot \nabla M_{\pm}(\mathbf{r},t)\vert_{\mathbf{r}\in \partial V} = 0$, which represents an ideal surface without spin depolarization effect. 
In the opposite limit $\lambda \rightarrow \infty$, the boundary condition becomes $M_{\pm}(\mathbf{r},t)|_{\mathbf{r} \in \partial V} = 0$, which means the spin magnetization is completely randomized at the wall.
Compared with the Eq.~(17) of Wu's review~\cite{Wu2021}, we have
\begin{equation}
  \lambda \approx \frac{3L}{4 \lambda_T} \xi_s^B, \quad {\rm when}~\xi_s^B \ll 1,
  \label{Eq:lambda_L_relation}
\end{equation}
where $\lambda_T$ is the mean free path of Xe atoms and $0<\xi_s^B \ll 1$ represents the depolarization probability of Xe spins on the wall. 
For noble gases the depolarization probability is very small, $\xi_s^B \lesssim 10^{-7}$.~\cite{Sheng2014} 
Since $\lambda_T$ and $\xi_s^B$ are physical parameters that should not depend on the size of $V$, the parameter $\lambda$ is actually linearly scales with $L$, i.e., $\lambda \propto L$. 
The typical value of $\lambda$ for RbH coated cells or uncoated Pyrex cells in an experimental Rb-Xe comagnetometer system is approximately $10^{-3} \sim 10^{-2}$ for $L \approx 1~{\rm cm}$ cubic cells. 
Figure~\ref{Fig:boundaryCond_measurement} shows an example of $\lambda$ measurement experiment. The magnitude of $\lambda$ is indeed small and has vast difference over different spin species. 
So, in the derivation below, the condition $\lambda \ll 1$ always holds, and we assume that $\lambda$ is strongly spin-species dependent. 

\subsection*{Perturbation Treatment}
\label{sec:perturbation_solution}
The general solution of Eq.~\eqref{Eq:TransverseEoM} has the form
\begin{equation}
  M_+({\bf r},t) = \sum_\alpha{a_\alpha {\rm e}^{-s_\alpha t} \Psi_\alpha({\bf r})},
\end{equation}
where $\{a_\alpha\}$ are expansion coefficients depending on initial state, $\{s_\alpha\}$ and $\{\Psi_\alpha({\bf r})\}$ are the spatial eigenvalues and eigenmodes of Eq.~\eqref{Eq:TransverseEoM}. When the nonuniform magnetic field $B_1(\mathbf{r})$ is small, due to the fast decay rate of excited modes, usually only one mode is experimentally observable (fast diffusion limit). Thus, we have
\begin{equation}
  M_+({\bf r},t) \approx a_0 {\rm e}^{-s_0 t} \Psi_0({\bf r}),
  \label{Eq:timeSolution_mode0}
\end{equation}
which describes the FID signal of Xe spins. 

The eigen equation of Eq.~\eqref{Eq:TransverseEoM} can write as
\begin{equation}
  \left[ \hat{H}_0 + \hat{H}_1 (\mathbf{r})  \right] \Psi_{\alpha} (\mathbf{r})
  = - s_{\alpha} \Psi_{\alpha} (\mathbf{r}),
  \label{Eq:eigenEq}
\end{equation}
where $\hat{H}_0 \equiv D \nabla^2 - (\rmi \gamma_{\rm Xe} B_0 +\Gamma_{\rm 2c} )$ and 
$\hat{H}_1 \equiv -\rmi \gamma_{\rm Xe} B_1 (\mathbf{r})$. 

Denote $\{\phi_{\alpha}(\bfr), \kappa_{\alpha}\}$ to be the eigen solution of $\hat{H}_0$ under the boundary condition Eq.~\eqref{Eq:BoundaryConditions}: 
\begin{equation}
  \begin{aligned}
  \nabla^2 \phi_{\alpha}(\bfr) =& -\kappa_{\alpha}^2 \phi_{\alpha}(\bfr). 
  \end{aligned}
  \label{Eq:def_kappa}
\end{equation}
Then, we can calculate the matrix form of $\hat{H}_0$ and $\hat{H}_1$ under the orthonormalized basis 
$\{ \phi_{\alpha} (\mathbf{r}) \}$ and apply the results of nondegenerate time-independent perturbation theory, only to remember that $\hat{H}_0$ and $\hat{H}_1$ are non-Hermitian. 
$H_0$ is diagonal, and the matrix elements of $H_1$ are $(H_1)_{\alpha \beta}= - \rmi \gamma_{\rm Xe} b_{\alpha \beta}$ 
with $b_{\alpha \beta}$ being the $B_1 (\mathbf{r})$ induced coupling between eigenmodes:
\begin{equation}
  b_{\alpha \beta} \equiv \int_V{\phi_\alpha (\mathbf{r}) B_1 (\mathbf{r}) 
                                      \phi_\beta (\mathbf{r}) {\rm d^3}\mathbf{r}}.
  \label{Eq:def_mode_averaged_magnetic_field}
\end{equation}
Above, the Greek indices $\alpha, \beta$ may contain multiple integer indices, e.g. $\alpha = [mnp]$. The fundamental mode in Eq.~\eqref{Eq:timeSolution_mode0}, which has the slowest decay rate, is denoted using $\alpha=0$ or $\beta=0$.

We are particularly interested in the perturbation correction of the eigenvalue 
$s_0$ of the fundamental mode, which can be directly observed via the FID 
frequency shift and the spin decay rate. The eigenvalue of the fundamental mode, up to third order correction, is
\begin{widetext}
\begin{equation}
\begin{aligned}
  s_0 &= \underbrace{D \kappa_0^2 + \Gamma_{\rm 2c} + \rmi \gamma_{\rm Xe} B_0}_{\rm 0^{th}\ order}
    + \underbrace{\rmi \gamma_{\rm Xe} b_{00}}_{\rm 1^{st}\ order}
    + \underbrace{\gamma_{\rm Xe}^2 \sum_{\alpha \neq 0}^{}{\frac{b_{0\alpha} b_{\alpha 0} }{D(\kappa_\alpha^2 - \kappa_0^2)}} }_{\rm 2^{nd}\ order}
      \underbrace{- \rmi \gamma_{\rm Xe}^3 \sum_{\alpha,\beta \neq 0}^{}{
                  \frac{b_{0\alpha}b_{\alpha\beta}b_{\beta0}}{D^2(\kappa_{\alpha}^2-\kappa_{0}^2)(\kappa_{\beta}^2-\kappa_{0}^2)}
         }}_{\rm 3^{rd}\ order\  (a)} 
         + \underbrace{ \rmi \gamma_{\rm Xe}^3 b_{00} \sum_{\alpha \neq 0}^{}{
                \frac{ b_{0\alpha} b_{\alpha 0} }{D^2(\kappa_{\alpha}^2-\kappa_{0}^2)^2}
                }}_{\rm 3^{rd}\ order \ (b)}
    \\ &\equiv s_0^{(0)} + s_0^{(1)} + s_0^{(2)} + s_0^{\rm (3)},
  \label{Eq:s0_pertb}
\end{aligned}
\end{equation}
\end{widetext}
where $s_0^{(k)}$ denotes the $k^{\rm th}$ order correction, and $\{ \kappa_\alpha^2 \}$ is the eigenvalues of Laplacian operator $\nabla^2$ as defined in Eq.~\eqref{Eq:def_kappa}. 
The FID signal of  Xe spins observed in experiment is usually determined by the fundamental mode eigenvalue $s_0$. The real part of $0^{\rm th}$ order correction represents the intrinsic relaxation rate of Xe spins, in which $\Gamma_{\rm 2c}$ is the relaxation due to gaseous-atom collisions (e.g. collision between Rb and Xe atoms), and $D \kappa_0^2$ is the relaxation due to wall interaction. 
The imaginary part of $0^{\rm th}$ order correction, $\rmi \gXe B_0$, is the main part of FID frequency that determined by the main field $B_0$. 

As the perturbation matrix element $b_{\alpha \beta}$ is real, the $\rm 2^{nd}$ order correction is real, which represents the extra relaxation rate caused by the nonuniform magnetic field. 
The $\rm 1^{st}$ and $\rm 3^{rd}$ order corrections are purely imaginary, which contribute to the frequency shift. 
The $b_{00}$ in $\rm 1^{st}$ order correction is a weighted average of the nonuniform field $B_1 (\mathbf{r})$ (along $z$ direction). In the absence of boundary relaxation ($\lambda=0$), the weight $\phi_0(\bfr)^2 = 1/V$ is constant over the space, making $b_{00}$ just a trivial average of $B_1 (\mathbf{r})$. 
For $0< \lambda \ll 1$, the boundary relaxation makes the fundamental mode $\phi_0(\bfr)$ slightly different from uniform distribution, and the first order correction will have a small dependence on $\lambda$. 

When the solution domain and $B_1(\bfr)$ have parity symmetry, the $\rm 1^{st}$ and $\rm 3^{rd}$ order corrections above may vanish due to the parity symmetry of eigenmodes $\{  \phi_\alpha(\bfr) \}$ (i.e. $b_{\alpha \beta} = 0$ when $\phi_\alpha(\bfr) B_1(\bfr) \phi_\beta(\bfr)$ is an odd function). Then, the effect of transverse magnetic field ($B_x, B_y$) is not negligible. 
Following the derivation in section "\nameref{sec:adx_pertb_3D_TorreyEq}" in the methods, we find that in order to calculate the effect of transverse magnetic field, we can simply replace all the $\left\{  b_{\alpha \beta} \right\}$ in Eq.~\eqref{Eq:s0_pertb} with the following $\left\{  b_{\alpha \beta}^{\rm (tot)} \right\}$: 
\begin{align}
  b^{(\rm tot)}_{\alpha\beta} \equiv & b_{\alpha \beta} + 
       \frac{\gXe}{
      2\left[ \gXe B_0 -\rmi \left( D \kappa_\beta^2 + \GXe{2c} \right) \right]
      } \sum_\gamma{ b^+_{\alpha\gamma} b^-_{\gamma \beta} }
   , 
   \label{Eq:def_btot} \\
   b^{\pm}_{\alpha \beta} \equiv & \int_V{\phi_\alpha(\bfr) \left[ 
    B_{x}(\bfr) \pm \rmi B_{y}(\bfr)
   \right] \phi_\beta(\bfr) \rmd^3 \bfr}. 
   \label{Eq:def_bpm}
\end{align}


For typical experiment conditions, $B_0$ is much larger than $B_1$, $B_x$, $B_y$ and $\left(D \kappa_0^2 + \GXe{2c} \right)/\gXe$. So, in Eq.~\eqref{Eq:def_btot}, compared with $B_1$, the effect of $B_x$ is suppressed by a factor of $B_x/B_0$ (and the same for $B_y$), i.e. $b_{\alpha \beta}^{(\rm tot)}$ is typically dominated by $b_{\alpha \beta}$. This validates the previous assumption Eq.~\eqref{Eq:assumptionOfBz}. The effect of transverse magnetic field gets important only when $b_{\alpha \beta}$, the contribution of $B_1$, vanishes due to symmetry reason (see \ref{sec:adx_odd_parity} for examples).

\subsection*{Frequency Shift from Gradient Field}
So far, we have not specified the shape of solution domain $V$. Thus, the solutions above are applicable to arbitrary $V$. Now, let us consider a cubic domain 
\begin{equation}
  V = \left\{ (x,y,z) \middle| -L/2 \le x,y,z \le +L/2 \right\},
  \label{Eq:def_cubic_damain}
\end{equation} 
with $L$ the cubic's side length. The eigenmodes in this cubic domain have a separable form (see section "\nameref{sec:adx_unperturbed_eigenmodes}" in the methods for details): 
\begin{equation}
  \phi_{mnp}(\bfr) = \phi_m(x)\phi_n(y)\phi_p(z), 
\end{equation} 
where
\begin{equation}
  \phi_p(z) = \frac{1}{\mathcal{N}_p} \sin{\left(\kappa_p z + \delta_p\right)}, \quad
  \delta_p = \frac{(p+1)\uppi}{2}. 
  \label{Eq:phiz_ansatz_maintext}
\end{equation}

Obviously, $\phi_p(z)$ has parity: $\phi_p(z)=\phi_p(-z)$ for even $p$ and $\phi_p(z)=-\phi_p(-z)$ for odd $p$. Using this symmetry, one can immediately derive that, for linear gradient field $B_1(\bfr) = G_1 \cdot z$, $b_{00}=0$ and $b_{0\alpha}b_{\alpha\beta}b_{\beta 0}=0, \forall \alpha,\beta \neq 0$. So, the frequency shift vanishes up to third order if we omit the effect of transverse magnetic field. 

To count the transverse field, we note that a linear gradient field with axial symmetry should have the form
\begin{equation}
  \mathbf{B} = G_1 \left[ - \frac{1}{2} \left( x \hat{x} + y \hat{y} \right) + z \hat{z} \right].
  \label{Eq:linear_gradient_3Ddistribution}
\end{equation}
Substitute this magnetic field distribution into Eqs.~\eqref{Eq:def_btot} and \eqref{Eq:def_bpm}, we get the first order correction of frequency shift in Eq.~\eqref{Eq:s0_pertb} to be (see section "\nameref{sec:adx_gradient_field_calculation}" in the methods for details)
\begin{equation}
  \delta \omega^{(1)}_{G_1} \equiv \rmi s_0^{(1)} 
     \approx -\frac{\gXe G_1^2 L^2}{48 B_0} \left( 1- \frac{2}{15} \lambda \right)
     + O(\lambda^2). 
  \label{Eq:freqShift_linG_1st}
\end{equation}
The $\lambda^0$ part of Eq.~\eqref{Eq:freqShift_linG_1st} has similar form with Eq.~\eqref{Eq:freq_shift_cates1988} by noting that $|\nabla B_x|^2 = |\nabla B_y|^2 = G_1^2/4$. 
Zheng {\it et al.} also obtained the $\lambda^0$ part of this formula for a cubic domain based on Redfield theory~\cite{Zheng2011}, and is consistent with our result (see \ref{sec:adx_cmp_literature} for details).

The second order correction, which contributes to relaxation rate, is
\begin{equation}
    \frac{1}{T_2^{G_1}} \equiv s_0^{(2)} 
      \approx \frac{\gamma_{\rm Xe}^2 G_1^2 L^4}{120 D} \left(  1 - \frac{\lambda}{3}  \right) + O(\lambda^2).
    \label{Eq:decayRate_linG_2nd}
\end{equation}
Equation~\eqref{Eq:decayRate_linG_2nd} is derived using $\{ b_{\alpha\beta} \}$, because $b_{0\alpha}b_{\alpha 0}$ does not vanish and the contribution from transverse field is negligible. 


For an axial symmetric, quadratic gradient field 
\begin{equation}
  \mathbf{B} = G_2 \left[ -  xz \hat{x} - yz \hat{y}  + z^2 \hat{z} \right], 
  \label{Eq:quad_gradient_3Ddistribution}
\end{equation}
parity symmetry can no more guarantee $b_{00}=0$ or $b_{0\alpha}b_{\alpha\beta}b_{\beta 0}=0, \forall \alpha,\beta \neq 0$. 
The frequency shifts and relaxation rate in Eq.~\eqref{Eq:s0_pertb} are (see section "\nameref{sec:adx_gradient_field_calculation}" in the methods for details)
\begin{align}
  \delta \omega^{(1)}_{G_2} \equiv \rmi s_0^{(1)} &\approx -\frac{\gamma_{\rm Xe} G_2 L^2 }{12} \left(   1 - \frac{2}{15} \lambda \right) + O(\lambda^2), 
  \label{Eq:freqShift_quadG_1st} \\
  \delta \omega^{(3)}_{G_2} \equiv \rmi s_0^{(3)} &\approx \frac{\gamma_{\rm Xe}^3 G_2^3 L^{10} }{D^2} \chi_1 \left(  1 + \chi_2 \lambda \right) + O(\lambda^2), 
  \label{Eq:freqShift_quadG_3rd} \\
  \frac{1}{T_2^{G_2}} \equiv s_0^{(2)} &\approx \frac{\gXe^2 G_2^2 L^6}{7560D} \left( 
    1 - \frac{2}{15} \lambda
   \right) + O(\lambda^2), 
   \label{Eq:decayRate_quadG_2nd} 
\end{align}
where 
\begin{align}
  \chi_1 \approx 6.68056 \times 10^{-8}, \quad 
  \chi_2 \approx 0.646886. 
\end{align}
Unlike the previously well known result Eq.~\eqref{Eq:freq_shift_cates1988}, these two frequency shifts are determined by the distribution of longitudinal field $B_1(\bfr)$ rather than transverse fields. They cannot be suppressed by a large $B_0$ and are sensitive to boundary relaxation. 

The substantial difference between Eqs.~\eqref{Eq:freqShift_linG_1st} and \eqref{Eq:freqShift_quadG_1st} comes from the symmetry of the magnetic field spatial distribution. 
First order correction is proportional to the $b^{(\rm tot)}_{00}$ defined in Eq.~\eqref{Eq:def_btot}, in which $b_{00} \sim B_1$ represents the contribution from longitudinal field, and the remaining part $\left( b^{(\rm tot)}_{00} - b_{00} \right) \sim \left( B_x^2 + B_y^2 \right)/B_0$ is the contribution from transverse field. 
In the calculation of Eq.~\eqref{Eq:freqShift_linG_1st}, we have $b_{00}=0$ due to the odd-parity of the linear gradient field. Thus, the frequency shift comes from the remaining part, which is proportional to $G_1^2 L^2$ and has a $B_0$ in the denominator. 
In the calculation of Eq.~\eqref{Eq:freqShift_quadG_1st}, $b_{00}$ is nonzero. Since $B_0$ is much larger than nonuniform fields $B_x$ and $B_y$, $b^{(\rm tot)}_{00}$ should be dominated by $b_{00}$. Thus, the frequency shift is proportional to $G_2 L^2$ and has no $B_0$ in the denominator. 


\subsection*{Systematic Error of Comagnetometer}
In NMR gyroscopes~\cite{Walker2016} and comagnetometer~\cite{Sheng2014} experiments, two different kinds of nuclear spins are used to compensate the fluctuation of main field $B_0$. However, Eqs.~\eqref{Eq:freqShift_linG_1st}, \eqref{Eq:freqShift_quadG_1st} and \eqref{Eq:freqShift_quadG_3rd} show that nonuniform magnetic field will lead to frequency shifts depending on $\lambda$, $\gamma$ and $D$, which can be different for different nuclear spins. These spin-species dependent frequency shifts may lead to imperfect compensation of magnetic field fluctuation. 
Following the analysis in section "\nameref{sec:adx_NMRG_sysError}" in the methods, Eqs.~\eqref{Eq:freqShift_linG_1st}, \eqref{Eq:freqShift_quadG_1st} and \eqref{Eq:freqShift_quadG_3rd} will respectly contribute systematic errors to the rotation signal of a NMR gyroscope as
\begin{align}
  \delta \Omega_{\rm rot}^{(G_1,1)} &\approx \frac{\lambda_{131} - \lambda_{129}}{L} 
         \frac{\bar{\gamma} G_1^2L^3}{360 B_0}   , 
  \label{Eq:deltaOmega_linG_1st} \\
  \delta \Omega_{\rm rot}^{(G_2,1)} &\approx \frac{\lambda_{131} - \lambda_{129}}{L} 
         \frac{\bar{\gamma} G_2 L^3}{90} , 
  \label{Eq:deltaOmega_quadG_1st} \\
  \delta \Omega_{\rm rot}^{(G_2,3)} &\approx - \chi_1 \left( \frac{\gamma_{129}^2}{D_{129}^2} - \frac{\gamma_{131}^2}{D_{131}^2} \right)
   \bar{\gamma} G_2^3 L^{10}, 
  \label{Eq:deltaOmega_quadG_3rd}
\end{align}
where $\delta \Omega_{\rm rot}^{(G_k,n)} = \Omega_{\rm rot}^{(2 \omega)} - \Omega_{\rm rot}$ is the difference between the gyroscope output $\Omega_{\rm rot}^{(2 \omega)}$ and the real rotation rate $\Omega_{\rm rot}$. 
 $\lambda_{129/131}$, $\gamma_{129/131}$ and $D_{129/131}$ are the boundary conditions, gyromagnetic ratios and diffusion constants for the \XeName{129} and \XeName{131} spins, respectively. The value of gyromagnetic ratios are~\cite{Makulski2015} $\gamma_{129} = \left. -2\uppi \left( 11.860156~{\rm mHz \cdot nT^{-1}} \right) \right.$, $\gamma_{131} = \left. +2\uppi \left(  3.515769~{\rm mHz \cdot nT^{-1}} \right) \right.$ and
\begin{equation}
  \begin{aligned}
\bar{\gamma} \equiv \frac{\gamma_{129} \gamma_{131}}{\gamma_{131}-\gamma_{129}} =& -2\uppi \left( 2.711874~{\rm mHz \cdot nT^{-1}} \right).
  \end{aligned}
\end{equation}
Systematic error of Eq.~\eqref{Eq:deltaOmega_quadG_3rd} was experimentally studied by Sheng {\it et al.} using $\rm ^{3}He$-\XeName{129} comagnetometer, showing a good consistency between experiment data and numerical simulation result~\cite{Sheng2014}. 

\begin{figure}
  \includegraphics[scale=1]{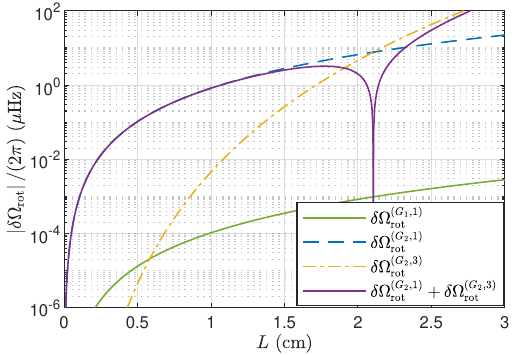}
  \caption[]{\label{Fig:compare_systematicErrors}{\textbf{The relative magnitude of different systematic errors.} 
  The systematic error from the third order frequency correction, $\delta \Omega_{\rm rot}^{(G_2,3)}$, is significant for large cell and negligible for small cell due to the strong dependence on the cell size $L$. 
  The characteristic cell length, where first and third order corrections have similar size, is in the order of 2~cm and inversely dependent on field gradient, as shown in Eq.~\eqref{Eq:characteristic_cell_length}. 
  The contribution from linear gradient field is much smaller than from quadratic field because of the suppression by a large main field $B_0$. 
  In the numerical calculation, $B_0=20000~{\rm nT}$, $G_1=10~{\rm nT \cdot cm^{-1}}$ and $G_2=10~{\rm nT \cdot cm^{-2}}$ are used. The values of $\lambda$ and $D$ are chosen to be the experimentally measured results presented in section "\nameref{sec:adx_wallRelax_measurement}" in the methods. 
  See Supplementary Data 1 for the source data of this figure.}}
\end{figure}

Figure~\ref{Fig:compare_systematicErrors} shows an example of the typical magnitude of these systematic errors for various cell size. 
For small cell, first order correction dominates the systematic error. Matching the $\lambda$ value of different spins helps reduce this systematic error. As cell size gets larger, third order (and higher order) correction become significant and finally blows up. 
The characteristic cell length where $\left| \delta \Omega_{\rm rot}^{(G_2,1)} \right| = \left| \delta \Omega_{\rm rot}^{(G_2,3)} \right|$ is 
\begin{equation}
L_{\rm c}^{(G_2)} = \left|  
  \frac{\lambda_{131} - \lambda_{129}}{90 \chi_1 G_2^2 L \left( \frac{\gamma_{129}^2}{D_{129}^2} - \frac{\gamma_{131}^2}{D_{131}^2} \right) }
\right|^{\frac{1}{7}}.
  \label{Eq:characteristic_cell_length}
\end{equation}

A 2~cm cubic cell (usually used in new physics searching experiments) can gain systematic errors in $10~{\rm \mu Hz}$ order. 
These errors should play an important role in understanding the fundamental precision limit of NMR gyroscopes and new physics detection based on comagnetometers. 
Our result can be used to explain the isotope shift effect observed in comagnetometer type experiments~\cite{{Bulatowicz2013,Vershovskii2018,Petrov2020,Vershovskii2020,Petrov2022}} by noticing that the alkali field has a highly nonuniform spatial distribution, leading to complex spin-species dependent frequency shifts.

Equations~\eqref{Eq:deltaOmega_linG_1st} and \eqref{Eq:deltaOmega_quadG_1st} convert the boundary condition $\lambda$ into a frequency signal $\delta \Omega_{\rm rot}^{(G_k,1)}$, thus can be used to measure the wall relaxation rate of nuclear spins. Compared to previous wall relaxation experiment, as shown in Fig.~\ref{Fig:boundaryCond_measurement}, which need to sweep cell temperature~\cite{Volk1980, Zeng1983, Nicol1984, Wu1990, Butscher1994, Driehuys1995, Wu2021}, our proposed method is much faster and capable of real-time monitoring the change of $\lambda$. 
The amplitude of $\delta \Omega_{\rm rot}^{(G_2,1)}/(2\uppi)$ in Eq.~\eqref{Eq:deltaOmega_quadG_1st} is approximately $5~{\rm \mu Hz}$ (using typical values $\lambda \approx 10^{-3}$ and $\bar{\gamma} G_2 L^2 / (2\uppi) \approx 0.5~{\rm Hz}$), and should be detectable using state-of-art comagnetometer technique. 
As a reference, the NMR gyroscope output signal reported in Fig.~9(a) of Gao's work~\cite{Gao2023} shows an $\sim 0.5~{\rm \mu Hz}$ RMS noise (after 100s average) and $\sim 2~{\rm \mu Hz}$ long time drift (over 8 hours). 
However, to experimentally apply this method, there still some challenges.
For example, the nonuniform alkali field generated by polarized alkali spins can interfere with the coil gradient field in the third order correction. A careful calibration of the effect of the alkali field is needed to ensure the accuracy of $\lambda$ measurement. The suppression of comagnetometer's long time drift is also important if real-time monitoring of $\lambda$ is desired.


\section*{Conclusion}
The frequency shift formulas presented in this work show quite different behaviors with the previously well known formula Eq.~\eqref{Eq:freq_shift_cates1988} for magnetic field gradient. It turns out that Eq.~\eqref{Eq:freq_shift_cates1988} is a special case where magnetic field distribution and solution domain both have parity symmetry. The use of Eq.~\eqref{Eq:freq_shift_cates1988} may severely underestimate the actual frequency shift caused by a nonuniform field. 

It is notable that different spins have slightly different frequency shift. 
This deviation of frequency shift can introduce significant systematic errors in comagnetometer experiments as well as other precision measurement experiments that rely on comparing the Larmor frequency of multiple kinds of spins. This systematic error could be one of the sources that limit the detection threshold of comagnetometer, which is a novel tool for dark matter searching~\cite{Yevgeny2017, Abel2017, Graham2018, Wu2019, Bloch2023}, exotic interaction detection~\cite{Tullney2013, Lee2018, Zhang2023a} and the verification of many other new physics models~\cite{Safronova2018}.  

Equations~\eqref{Eq:deltaOmega_linG_1st} and \eqref{Eq:deltaOmega_quadG_1st} provide a tool for the study of spin-wall interaction. Since frequency measurement is one of the most precise measurements, this method has great advantages in precision and bandwidth. It should be a tool with great potential for spin-solid interaction research.

%

%

\section*{Methods}
(Note: In order not to introduce confusion, in this article, subscript $m, n, p$ will always represent a single integer index, while a Greek subscript such as $\alpha, \beta, \gamma$ represents multiple indices.)

\subsection*{Perturbation Treatment of the 3D Torrey Equation}
\label{sec:adx_pertb_3D_TorreyEq}
This section considers the effect of transverse magnetic field in the presence of a large main field $B_0$. 
Equation~\eqref{Eq:BlochEquation} can rewrite to the following form: \linebreak[4]
\begin{align}
  &\begin{aligned}
      \frac{\partial M_z}{\partial t} =& D \nabla^2 M_z 
           - \GXe{1c} M_z + R_{\rm p}' S_z \\
           &- \frac{\rmi}{2} \gXe \left( B_+ M_- - B_- M_+ \right), 
  \end{aligned}
  \label{Eq:adx_TorreyEq_seperateForm_Mz} \\
  &\begin{aligned}
      \frac{\partial M_+}{\partial t} =& D \nabla^2 M_+ 
           - \rmi \gXe \left( B_z M_+ - B_+ M_z \right) - \GXe{2c} M_+, 
  \end{aligned}
  \label{Eq:adx_TorreyEq_seperateForm_M+}
\end{align}
with
\begin{align}
  \mathbf{B} \equiv&  B_x(\bfr) \hat{x} + B_y(\bfr) \hat{y} + B_z(\bfr) \hat{z}, \\
  B_\pm \equiv& B_x(\bfr) \pm \rmi B_y(\bfr), 
\end{align}
and the boundary condition
\begin{equation}
  \left.\left(\mathbf{n}\cdot \nabla M_{i}(\mathbf{r},t) + \frac{\lambda}{L}\cdot M_{i}(\mathbf{r},t) \right) \right\vert_{\mathbf{r} \in \partial V} = 0,
  \label{Eq:adx_BoundaryConditions}
\end{equation}
where $M_i$ stands for $M_x, M_y$ or $M_z$. $\GXe{1c} \equiv \GXe{10}+R_{\rm p}', \GXe{2c} \equiv \GXe{20}+R_{\rm p}'$. 
Suppose $B_z = B_0 + B_1(\bfr)$ where $B_0$ is a large uniform main field and $B_1(\bfr)$ is a small nonuniform field. Introduce the rotating frame with
$\tilde{M}_\pm \equiv M_\pm~\rme^{\pm \rmi \gXe B_0 t}$. Then, Eqs.~\eqref{Eq:adx_TorreyEq_seperateForm_Mz} and \eqref{Eq:adx_TorreyEq_seperateForm_M+} becomes
\begin{align}
  &\begin{aligned}
      \frac{\partial M_z}{\partial t} =& D \nabla^2 M_z 
           - \GXe{1c} M_z + R_{\rm p}' S_z \\
           &- \frac{\rmi}{2} \gXe \left( B_+ \tilde{M}_- \rme^{+\rmi \gXe B_0 t} - B_- \tilde{M}_+ \rme^{-\rmi \gXe B_0 t} \right),  
  \end{aligned}
  \label{Eq:adx_TorreyEq_seperateForm2_Mz}\\
  &\begin{aligned}
      \frac{\partial \tilde{M}_+}{\partial t} =& D \nabla^2 \tilde{M}_+ - \GXe{2c} \tilde{M}_+ \\
           &- \rmi \gXe B_1 \tilde{M}_+ +  \rmi \gXe B_+ M_z \rme^{+\rmi \gXe B_0 t}  .     
  \end{aligned}
  \label{Eq:adx_TorreyEq_seperateForm2_M+}
\end{align}
In the rotating frame, the change of $\tilde{M}_{\pm}$ and $M_{z}$ should be slow compared to Larmor frequency $\gXe B_0$. So, in Eqs.~\eqref{Eq:adx_TorreyEq_seperateForm2_Mz} and \eqref{Eq:adx_TorreyEq_seperateForm2_M+}, the high frequency terms which contain $\rme^{\pm \rmi \gXe B_0 t}$ factor can be directly ignored to the first approximation. This is called the RWA, leading to the decoupling of $M_z$ and $\tilde{M}_\pm$, and justifies the previous assumption of omitting the transverse component of magnetic field (Eq.~\eqref{Eq:assumptionOfBz}). 

However, according to Eq.~\eqref{Eq:adx_TorreyEq_seperateForm2_Mz}, the solution of $M_z$ can contain a small $\rme^{- \rmi \gXe B_0 t}$ component. Then, the $M_z \rme^{+\rmi \gXe B_0 t}$ term in Eq.~\eqref{Eq:adx_TorreyEq_seperateForm2_M+} will generate a DC contribution which might have some effects to the solution of $\tilde{M}_+$. 
So, let us consider the case when RWA is not directly applied to Eqs.~\eqref{Eq:adx_TorreyEq_seperateForm2_Mz} and \eqref{Eq:adx_TorreyEq_seperateForm2_M+}. 

Denote the eigenmodes of $\nabla^2$ operator under the above boundary condition as
\begin{equation}
  \nabla^2 \phi_\alpha(\bfr) = -\kappa_\alpha^2 \phi_\alpha(\bfr), \quad
  \int_V{\phi_\alpha(\bfr) \phi_\beta(\bfr) \rmd^3 \bfr} = \delta_{\alpha \beta}.
  \label{Eq:adx_def_eigenmode}
\end{equation}
Then, the solution of $\mathbf{M}$ can be expanded as
\begin{equation}
      M_z \equiv  \sum_\alpha{c_{z,\alpha}(t) \phi_\alpha(\mathbf{r})}, \quad 
      \tilde{M}_\pm \equiv \sum_\alpha{\tilde{c}_{\pm, \alpha}(t) \phi_\alpha(\mathbf{r})}, 
  \label{Eq:adx_eigenmode_expansion_of_M_Sz}
\end{equation}
with $\tilde{c}_{-,\alpha} = \tilde{c}^*_{+,\alpha}$.
Using these expansions and the orthonormality of eigenmodes, one can transform Eqs.~\eqref{Eq:adx_TorreyEq_seperateForm2_Mz} and \eqref{Eq:adx_TorreyEq_seperateForm2_M+} into the following linear equation system of expansion coefficients:
\begin{align}
    &\begin{aligned}
      \frac{\rmd c_{z,\alpha}}{dt} =& -\left(D \kappa_\alpha^2 + \GXe{1c}\right) c_{z,\alpha} + R_{\rm p}' d_{z,\alpha} \\
      & - \frac{\rmi}{2} \gXe
        \rme^{+ \rmi \gXe B_0 t} \sum_\beta{ b^+_{\alpha \beta} \tilde{c}_{-,\beta}} \\
      & + \frac{\rmi}{2} \gXe
        \rme^{- \rmi \gXe B_0 t} \sum_\beta{ b^-_{\alpha \beta} \tilde{c}_{+,\beta}},
     \end{aligned}
     \label{Eq:adx_TorreyEq_coeffsForm_cz} \\
    &\begin{aligned}
      \frac{\rmd \tilde{c}_{+,\alpha}}{dt} =& 
      -\left(D \kappa_\alpha^2 + \GXe{2c}\right) \tilde{c}_{+,\alpha} \\
      &- \rmi \gXe \sum_\beta{b_{\alpha \beta} \tilde{c}_{+,\beta}}
      + \rmi \gXe \rme^{+\rmi \gXe B_0 t} \sum_\beta{b^+_{\alpha \beta} c_{z,\beta}}.     
  \end{aligned}
  \label{Eq:adx_TorreyEq_coeffsForm_c+}
\end{align}
where
\begin{align}
    b^{\pm}_{\alpha \beta} \equiv& \int_V{\phi_\alpha(\bfr) B_{\pm}(\bfr) \phi_\beta(\bfr) \rmd^3 \bfr}, \label{Eq:adx_def_bMatrix1} \\
    b_{\alpha \beta} \equiv& \int_V{\phi_\alpha(\bfr) B_1(\bfr) \phi_\beta(\bfr) \rmd^3 \bfr}, \label{Eq:adx_def_bMatrix2} \\
    d_{z,\alpha} \equiv& \int_V{\phi_\alpha(\bfr) S_z(\bfr) \rmd^3 \bfr}.
  \label{Eq:adx_def_bMatrix3}
\end{align}
Equations~\eqref{Eq:adx_TorreyEq_coeffsForm_cz} and \eqref{Eq:adx_TorreyEq_coeffsForm_c+} can be directly used in numerical simulation. 

Integrate the $c_{z,\alpha}$ equation in Eq.~\eqref{Eq:adx_TorreyEq_coeffsForm_cz}, one gets
\begin{equation}
  \begin{aligned}
      c_{z,\alpha}(t) =& c_{z,\alpha}(0) + \int_0^t{
        \left[ R_{\rm p}' d_{z,\alpha} - \left( D \kappa_\alpha^2 + \GXe{1c} \right) c_{z,\alpha}(t')  \right] dt'
      } \\ 
      &- \frac{\rmi}{2} \gXe \sum_\beta{
          b_{\alpha\beta}^+ \left(\int_0^t{ \rme^{+\rmi \gXe B_0 t'} \tilde{c}_{-,\beta}(t') dt' }\right)} \\
      &+ \frac{\rmi}{2} \gXe \sum_\beta{
           b_{\alpha\beta}^- \left(\int_0^t{ \rme^{-\rmi \gXe B_0 t'} \tilde{c}_{+,\beta}(t') dt' }\right)
      }.
  \end{aligned}
  \label{Eq:adx_int_czalpha}
\end{equation}
Based on the picture of Larmor precession and numerical simulation, we can safely assume that 
\begin{equation}
  \tilde{c}_{\pm, \alpha}(t) \approx \tilde{c}_{\pm, \alpha}(0) \exp{\left( \mp \rmi \omega_{\alpha}t - \GXe{\alpha}t \right)},
\end{equation}
where $\GXe{\alpha} \approx D \kappa_\alpha^2 + \GXe{2c}$ is the relaxation rate of $\phi_\alpha(\bfr)$ mode, and $\omega_{\alpha}$ is the frequency correction of this mode. Then, the integration above becomes
\begin{equation}
  \begin{aligned}
  &\int_0^t{ \rme^{ \pm \rmi \gXe B_0 t'} \tilde{c}_{\mp,\beta}(t') dt' } \\
   =& \frac{c_{\mp,\beta}(0)}{\pm \rmi \left( \gXe B_0 + \omega_\beta \right) - \GXe{\beta}}
    \left[ \rme^{\pm \rmi \left( \gXe B_0 + \omega_\beta \right)t - \GXe{\beta}t} - 1  \right].
  \label{Eq:adx_int_czalpha2}
  \end{aligned}
\end{equation}

Insert Eqs.~\eqref{Eq:adx_int_czalpha} and \eqref{Eq:adx_int_czalpha2} into the $\tilde{c}_{+,\alpha}$ equation of Eq.~\eqref{Eq:adx_TorreyEq_coeffsForm_c+}. Noticing that $|\omega_\beta| \ll |\gXe B_0|$, we can use RWA, ignoring all the terms with $\rme^{+\rmi \gXe B_0 t}$ or $\rme^{+\rmi \left( 2\gXe B_0 + \omega_\beta \right) t}$ factor. 
Finally, we get
\begin{equation}
  \begin{aligned}
    \frac{\rmd \tilde{c}_{+,\alpha}}{dt} \approx& 
    -\left(D \kappa_\alpha^2 + \GXe{2c}\right) \tilde{c}_{+,\alpha}
    - \rmi \gXe \sum_\beta{b_{\alpha \beta} \tilde{c}_{+,\beta}}\\
    &+ \frac{\gXe^2}{2}  \sum_{\beta,\gamma}{\frac{b^+_{\alpha \gamma} b^-_{\gamma\beta} \tilde{c}_{+,\beta}}{\GXe{\beta} + \rmi \left( \gXe B_0 + \omega_\beta \right)}} \\
    \approx& 
    -\left(D \kappa_\alpha^2 + \GXe{2c}\right) \tilde{c}_{+,\alpha}
    - \rmi \gXe \sum_\beta{b^{(\rm tot)}_{\alpha \beta} \tilde{c}_{+,\beta}}, 
\end{aligned}
\label{Eq:adx_TorreyEq_coeffsForm_transverse}
\end{equation}
where
\begin{equation}
  b^{(\rm tot)}_{\alpha\beta} \equiv b_{\alpha\beta} + \frac{\gXe}{
    2\left[ \gXe B_0 -\rmi \left( D \kappa_\beta^2 + \GXe{2c} \right) \right]
    } \sum_\gamma{
    b^+_{\alpha\gamma} b^-_{\gamma \beta}
  }.
  \label{Eq:adx_balphabeta_tot_formula}
\end{equation}

If we directly apply RWA to Eq.~\eqref{Eq:adx_TorreyEq_seperateForm2_M+}, which means ignoring all the effects of $B_x$ and $B_y$, then Eq.~\eqref{Eq:adx_TorreyEq_coeffsForm_transverse} becomes
\begin{equation}
    \frac{\rmd \tilde{c}_{+,\alpha}}{dt} 
    \approx
    -\left(D \kappa_\alpha^2 + \GXe{2c}\right) \tilde{c}_{+,\alpha}
    - \rmi \gXe \sum_\beta{b_{\alpha \beta} \tilde{c}_{+,\beta}}, 
\label{Eq:adx_TorreyEq_coeffsForm_transverse_RWA}
\end{equation}
which is equivalent to the eigen equation Eq.~\eqref{Eq:eigenEq}. 
Comparing Eqs.~\eqref{Eq:adx_TorreyEq_coeffsForm_transverse} and \eqref{Eq:adx_TorreyEq_coeffsForm_transverse_RWA}, it is easy to see that, to account the leading order effect of $B_x$ and $B_y$, we just need to replace the $\{b_{\alpha \beta}\}$ in Eq.~\eqref{Eq:s0_pertb} with $\{b^{(\rm tot)}_{\alpha \beta} \}$.  
Since the main field $B_0$ is much larger than nonuniform field $B_1$, $B_x$ and $B_y$, $b_{\alpha\beta}^{(\rm tot)}$ is dominated by $b_{\alpha \beta}$. The effect of $B_{x/y}$ is suppressed by a factor $B_{x/y} / B_0$ compared with $B_1$. 
Also, the imaginary part of $b_{\alpha\beta}^{(\rm tot)}$ should be much smaller than its real part.

\subsection*{Unperturbed Eigenmodes in Cubic Domain}
\label{sec:adx_unperturbed_eigenmodes}
This section will derive the eigenmodes and eigenvalues of the Torrey equation in a cubic domain of the form Eq.~\eqref{Eq:def_cubic_damain}.

Consider the following eigen equation: 
\begin{equation}
    D \nabla^2 \phi_{mnp} - (\rmi \gamma_{\rm Xe} B_0 + \Gamma_{\rm 2c}) \phi_{mnp}
    = - s_{mnp}^{(0)} \phi_{mnp},
    \label{Eq:adx_eigenEq_const}
  \end{equation}
where $-s_{mnp}^{(0)}$ is the eigenvalue of the eigenmode $\phi_{mnp}(\mathbf{r})$. 

The eigenmodes $\{ \phi_{mnp}(\mathbf{r}) \}$ of Eq.~\eqref{Eq:adx_eigenEq_const} 
can be written in a factorized form of
\begin{equation}
  \phi_{mnp}(\mathbf{r}) = \phi_m(x)\phi_n(y)\phi_p(z),
  \label{Eq:adx_eigenMode_sepForm}
\end{equation}
where $m$, $n$ and $p$ are non-negative integers labelling the eigenmodes 
in the $x$, $y$ and $z$ directions, respectively. 
It is easy to check that the 1D eigenmodes have the form
\begin{equation}
  \phi_p(z) = \frac{1}{\mathcal{N}_p} \sin{\left(\kappa_p z + \delta_p\right)},
  \label{Eq:adx_phiz_ansatz}
\end{equation}
where $\mathcal{N}_p$ is normalization factor, and $\kappa_p$ and $\delta_p$ are 
real numbers determined by the boundary condition Eq.~\eqref{Eq:BoundaryConditions}. 
The wave numbers $\kappa_p$ are the solutions of the transcendental equation
\begin{equation}
  \tan{\left(\kappa_p L\right)} = \frac{2 \lambda \kappa_p L}{\kappa_p^2 L^2 - \lambda^2},
  \label{Eq:adx_kappa_eq}
\end{equation}
and the phase shifts are determined by
\begin{equation}
  \tan{\left(\delta_p - \frac{\kappa_p L}{2}\right)} = \frac{\kappa_p L}{\lambda}.
  \label{Eq:adx_delta_eq}
\end{equation}
Using the normalization condition $\int{\phi_p^2(z) {\rm d}z} = 1$, 
the normalization factor is
\begin{equation}
  \mathcal{N}_p = \sqrt{ \frac{L}{2} + \frac{\lambda L}{\kappa_p^2 L^2 + \lambda^2} }.
  \label{Eq:adx_normalizationFactor}
\end{equation}
The eigenvalues corresponding to $\phi_{mnp}(\mathbf{r})$ are 
\begin{equation}
  s^{(0)}_{mnp} = D \kappa_{mnp}^2 + \Gamma_{\rm 2c} + \rmi \gamma_{\rm Xe} B_0,
  \label{Eq:adx_eigenvalue_0th}
\end{equation}
where $\kappa_{mnp}^2 \equiv \kappa_m^2+\kappa_n^2+\kappa_p^2$. The real part of the 
eigenvalue, $\Gamma_2=D \kappa_{mnp}^2+\Gamma_{\rm 2c}$, is the decay rate of the 
eigenmode $\phi_{mnp}(\mathbf{r})$, and the imaginary part, $\gamma_{\rm Xe} B_0$, is the spin 
precession frequency in the uniform magnetic field $B_0$. The superscript of 
$s_{mnp}^{(0)}$ represents that this is the $\rm 0^{th}$ order correction of the perturbation solution presented in Eq.~\eqref{Eq:s0_pertb}.

\begin{figure}
  \includegraphics[scale=1]{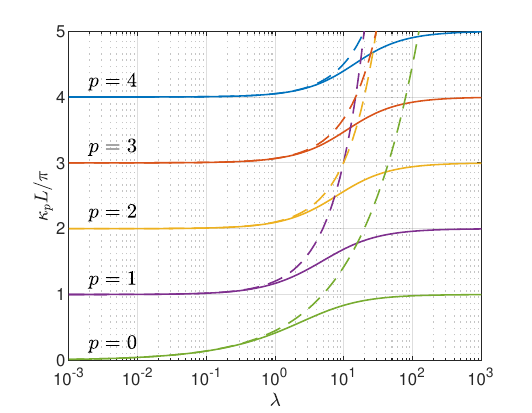}
  \caption[]{\label{Fig:adx_eigenvalue_num}{\textbf{The eigen spectrum of spatial eigenmodes for different boundary relaxation strength.} 
             The solid lines are the exact value calculated from numerical method. 
             The dashed lines are the approximation value calculated from Eq.~\eqref{Eq:adx_kappa_solution}.
             See Supplementary Data 1 for the source data of this figure.}}
\end{figure}

Figure~\ref{Fig:adx_eigenvalue_num} gives the numerical solutions of $\kappa_p$ for various $\lambda$ values. 
To reveal the underlying physics of the wall-relaxation, we expand the tangent 
function in Eq.~\eqref{Eq:adx_kappa_eq} in the neighborhood of $\kappa_p L=p \uppi$, 
and find the solution of wave number in the limit of $\lambda \ll 1$ to be
\begin{align}
  \kappa_p \approx
  \left\{ 
    \begin{aligned}
      & \frac{\sqrt{2\lambda}}{L}                                           &,& p=0 \\
      & \frac{p \uppi}{L} + \frac{2 \lambda}{p \uppi L} \approx \frac{p \uppi}{L} &,& p=1,2,\dots
    \end{aligned} 
  \right. .
  \label{Eq:adx_kappa_solution}
\end{align}
The exact solution of $\delta_p$ (without assuming $\lambda \ll 1$) is 
\begin{align}
    \delta_p = \frac{(p+1)\uppi}{2}.
    \label{Eq:adx_delta_solution}
\end{align}
For the fundamental mode ($m=n=p=0$), the wall-interaction contributes a relaxation rate through diffusion: 
\begin{equation}
  \Gamma_{000} \equiv D \kappa_{000}^2 \approx 6 \lambda \frac{D}{L^2}.
\end{equation}
For the excited modes ($[m,n,p] \neq [0,0,0]$), the relaxation rate due to spin 
diffusion is
\begin{equation}
  \Gamma_{mnp} \equiv D \kappa_{mnp}^2 \approx (m^2+n^2+p^2) \uppi^2 \frac{D}{L^2}.
\end{equation}
As $\lambda \lesssim 10^{-2}$, even for the lowest excited 
modes (with $m^2+n^2+p^2=1$), the decay rate is much faster than the fundamental mode, 
i.e., $\Gamma_{mnp} \gg \Gamma_{000}$. 

In the $\lambda \ll 1$ limit, $\phi_p(z)$ are approximately 
\begin{align}
&\begin{aligned}
  \phi_0(z) &\approx \frac{1 - \lambda z^2 / L^2}{\sqrt{L(1 - \lambda/6 + \lambda^2/80)}}, 
  \end{aligned}
  \label{Eq:adx_phiz_approx_0} \\
&\begin{aligned}
  \phi_p(z) &\approx 
  \left\{    
  \begin{matrix}
      \frac{1}{\mathcal{N}_p} \sin{\left[  \left( \frac{p \uppi}{L} + \frac{2 \lambda}{p \uppi L} \right) z \right]},  &  p~{\rm is~odd}  \\
      \frac{1}{\mathcal{N}_p} \cos{\left[  \left( \frac{p \uppi}{L} + \frac{2 \lambda}{p \uppi L} \right) z \right]},  &  p~{\rm is~even}  \\
  \end{matrix}
       \right., p > 0.
\end{aligned}
  \label{Eq:adx_phiz_approx_p}
\end{align}


\subsection*{Frequency Shift Formulas for Gradient Fields}
\label{sec:adx_gradient_field_calculation}
In this section, we calculate the frequency shift formulas for two simple forms of nonuniform magnetic field. 

\subsubsection*{Linear Gradient Field}
Let us first consider a linear gradient field $B_z = B_0 +  \left. G_1 \cdot z \right.$. 
Since a real magnetic field should obey the Gauss law, if we assume axial symmetry, magnetic field should have the following distribution:
\begin{equation}
  \mathbf{B}(\bfr) = G_1 \left[ - \frac{1}{2} \left( x \hat{x} + y \hat{y} \right) + z \hat{z} \right] + B_0 \hat{z}, 
  \quad B_1(\bfr) = G_1 z. 
\end{equation}
This is the field distribution of linear gradient compensation coil normally used in NMR experiments. 

The first order term $b_{00}$ is zero, for the reason that $B_1(\bfr) = G_1 z$ is odd function and $\phi_0(z)$ in Eq.~\eqref{Eq:adx_phiz_approx_0} is even function. So, the first order frequency shift of a linear gradient field is mainly contributed from transverse components. We need to calculate the $b_{00}^{\rm (tot)}$ in Eq.~\eqref{Eq:adx_balphabeta_tot_formula}. Let us first calculate $b^{\pm}_{0 \alpha}$ using the definition Eq.~\eqref{Eq:adx_def_bMatrix1}. According to the parity of the approximate eigenmode expressions in Eqs.~\eqref{Eq:adx_phiz_approx_0} and \eqref{Eq:adx_phiz_approx_p}, $b^{\pm}_{0 \alpha}$ is nonzero only when $\alpha = [2p-1,0,0]$ or $\alpha = [0, 2p-1, 0]$, $p~\ge~1$. Thus, we have
\begin{equation}
  \begin{aligned}
     &b^{\pm}_{0,[2p-1,0,0]} = - \frac{G_1}{2} \int_{-L/2}^{+L/2}{ x \phi_0(x) \phi_{2p-1}(x) \rmd x } \\
    \approx&  (-1)^{p} \frac{\sqrt{2} G_1 L}{(2p-1)^2 \uppi^2} \left\{ 
        1 - \left[ \frac{1}{6} - \frac{1}{(2p-1)^2 \uppi^2} \right] \lambda
     \right\} \\
     &+ O(\lambda^2),
  \end{aligned}
\end{equation}
\begin{equation}
  \begin{aligned}
     &b^{\pm}_{0,[0,2p-1,0]} = - \frac{G_1}{2} \int_{-L/2}^{+L/2}{ (\pm \rmi y) \phi_0(y) \phi_{2p-1}(y) \rmd y } \\
    \approx&  \pm \rmi (-1)^{p} \frac{\sqrt{2} G_1 L}{(2p-1)^2 \uppi^2} \left\{ 
        1 - \left[ \frac{1}{6} - \frac{1}{(2p-1)^2 \uppi^2} \right] \lambda
     \right\} \\
     &+ O(\lambda^2).
  \end{aligned}
\end{equation}

Above, we use Eqs.~\eqref{Eq:adx_phiz_approx_0} and \eqref{Eq:adx_phiz_approx_p} as the approximate expressions of eigenmodes, and expand the result around $\lambda = 0$. Combining them together, we have 
\begin{equation}
  \begin{aligned}
    &\sum_{\alpha}{b^+_{0 \alpha} b^-_{\alpha 0}} \\
    = & \sum_{p=1}^{\infty}{ \left(  
          b^+_{0,[2p-1,0,0]} b^-_{0,[2p-1,0,0]}
         +b^+_{0,[0,2p-1,0]} b^-_{0,[0,2p-1,0]}   \right) } \\
    \approx & 2 \sum_{p=1}^{\infty}{\left(
      \frac{\sqrt{2} G_1 L}{(2p-1)^2 \uppi^2} \left\{ 
        1 - \left[ \frac{1}{6} - \frac{1}{(2p-1)^2 \uppi^2} \right] \lambda
     \right\}\right)^2
    } \\
    \approx & \frac{G_1^2 L^2}{24} \left( 1 - \frac{2}{15} \lambda \right) + O(\lambda^2).
  \end{aligned}
\end{equation}

The first order frequency shift is:
\begin{equation}
  \begin{aligned}
    \rmi s_0^{(1)} =& -\gXe b_{00}^{\rm (tot)} \approx -\gXe \frac{ \sum_{\alpha}{b^+_{0 \alpha} b^-_{\alpha 0}}  }{2B_0} \\
    \approx& -\frac{\gXe G_1^2 L^2}{48 B_0} \left( 1 - \frac{2}{15} \lambda \right) + O(\lambda^2).
  \end{aligned}
\end{equation}

Second order correction mainly consists of the contribution from longitudinal component. Due to symmetry reason, $b_{0\alpha}$ is nonzero only when $\alpha = [0,0,2p-1], p \ge 1$. Using the approximate eigenmode in Eqs.~\eqref{Eq:adx_phiz_approx_0} and \eqref{Eq:adx_phiz_approx_p}, we have
\begin{equation}
  \begin{aligned}
      &b_{0,[0,0,2p-1]} = G_1 \int_{-L/2}^{+L/2}{ z \phi_0(z) \phi_{2p-1}(z) \rmd z } \\
      \approx & (-1)^{p+1} \frac{2\sqrt{2} G_1 L}{(2p-1)^2 \uppi^2} \left\{ 
          1 - \left[ \frac{1}{6} - \frac{1}{(2p-1)^2 \uppi^2} \right] \lambda
       \right\} \\
       &+ O(\lambda^2).
  \end{aligned}
\end{equation}
The result above is a Taylor expansion around $\lambda=0$. 
So, the second order correction is
\begin{equation}
  \begin{aligned}
      s_0^{(2)} =& \frac{\gXe^2 L^2}{D} \sum_{p=1}^{\infty}{
        \frac{\left( b_{0,[0,0,2p-1]} \right)^2}{\left[ (2p-1)\uppi + \frac{2\lambda}{(2p-1)\uppi} \right]^2 - 2 \lambda}  
      } \\
      \approx& \frac{\gXe^2 L^2}{D} \sum_{p=1}^{\infty}{
        \frac{8 G_1^2 L^2 }{(2p-1)^6 \uppi^6} \left( 1 - \frac{\lambda}{3} \right) 
      } \\
      =& \frac{\gXe^2 G_1^2 L^4}{120 D} \left( 1 - \frac{\lambda}{3} \right) + O(\lambda^2).
  \end{aligned}
\end{equation}


\subsubsection*{Quadratic Gradient Field}
For quadratic gradient field $B_z = B_0 +  G_2 \cdot z^2$, if assume axial symmetry, the spatial distribution should be
\begin{equation}
  \mathbf{B} = G_2 \left[ -  xz \hat{x} - yz \hat{y}  + z^2 \hat{z} \right] + B_0 \hat{z}, 
  \quad B_1(\bfr) = G_2 z^2. 
\end{equation}
First order correction can be calculated directly using the approximate eigenmode in Eq.~\eqref{Eq:adx_phiz_approx_0}:
\begin{equation}
  \begin{aligned}
    b_{00} =& G_2 \int_{-\frac{L}{2}}^{+\frac{L}{2}}{z^2 \phi_0(z) \phi_0(z) \rmd z} \\
    \approx& \frac{G_2 L^2}{12} \left( 1 - \frac{2}{15} \lambda \right) + O(\lambda^2). 
    \label{Eq:adx_quadG_b00}
  \end{aligned}
\end{equation}

The calculation of second order correction is similar to linear gradient field. Due to symmetry reason, $b_{0\alpha}$ is nonzero only when $\alpha = [0,0,2p], p \ge 1$. Using the approximate eigenmode in Eqs.~\eqref{Eq:adx_phiz_approx_0} and \eqref{Eq:adx_phiz_approx_p}, we have
\begin{equation}
  \begin{aligned}
      &b_{0,[0,0,2p]} = G_2 \int_{-L/2}^{+L/2}{ z^2 \phi_0(z) \phi_{2p}(z) \rmd z } \\
      \approx & (-1)^{p} \frac{G_2 L^2}{\sqrt{2} p^2 \uppi^2} \left[ 
          1 - \left( \frac{1}{6} - \frac{5}{4 p^2 \uppi^2} \right) \lambda
       \right] + O(\lambda^2).
       \label{Eq:adx_quadG_b0alpha}
  \end{aligned}
\end{equation}
Thus, the second order correction is
\begin{equation}
  \begin{aligned}
      s_0^{(2)} =& \frac{\gXe^2 L^2}{D} \sum_{p=1}^{\infty}{
        \frac{\left( b_{0,[0,0,2p]} \right)^2}{\left[ 2p \uppi + \frac{2\lambda}{2p \uppi} \right]^2 - 2 \lambda}  
      } \\
      \approx& \frac{\gXe^2 L^2}{D} \sum_{p=1}^{\infty}{
        \frac{G_2^2 L^4 }{8 p^6 \uppi^6} \left[ 1 - \left( \frac{1}{3} - \frac{2}{p^2 \uppi^2} \right) \lambda \right] 
      } \\
      =& \frac{\gXe^2 G_2^2 L^6}{7560 D} \left( 1 - \frac{2}{15} \lambda \right) + O(\lambda^2).
  \end{aligned}
\end{equation}

The calculation of third order correction is a bit complicated. The calculation of $s_0^{\rm (3b)}$ term is similar to $s_0^{(2)}$:
\begin{equation}
  \begin{aligned}
      s_0^{\rm (3b)} \approx& \rmi \frac{\gXe^3}{D^2} L^4 b_{00} \sum_{p=1}^{\infty}{
          \frac{\left( b_{0,[0,0,2p]} \right)^2}{
              \left[ \left( 2p\uppi + \frac{2\lambda}{2p\uppi}  \right)^2 -2\lambda \right]^2
          }
      }\\
      \approx & \rmi \frac{\gXe^3 G_2^3 L^{10}}{3628800 D^2} \left( 1 - \frac{52}{165} \lambda \right) + O(\lambda^2).
  \end{aligned}
  \label{Eq:adx_quadG_s03b}
\end{equation}

When calculating $s_0^{\rm (3a)}$, we need the matrix element $b_{\alpha \beta}$. Due to symmetry reason, only when $\left. \alpha = [0,0,2p]\right.$ and $\left. \beta=[0,0,2n]\right.$, the product $b_{0\alpha} b_{\alpha\beta} b_{\beta 0}$ in the numerator of $s_0^{\rm (3a)}$ is nonzero. Using the approximate eigenmode in Eqs.~\eqref{Eq:adx_phiz_approx_0} and \eqref{Eq:adx_phiz_approx_p}, we have
\begin{equation}
    \begin{aligned}
        b_{\alpha\beta} = & G_2 \int_{-L/2}^{+L/2}{z^2 \phi_{2p}(z) \phi_{2n}(z) \rmd z} \\
        \approx &
        \begin{cases}
            \frac{(-1)^{n+p}G_2 L^2 (n^2+p^2)}{(n^2-p^2)^2 \uppi^2} 
            \left[ 1 - \frac{n^4-10n^2p^2+p^4}{4 n^2p^2(n^2+p^2)\uppi^2} \lambda \right],&  p \neq n  \\
            \frac{G_2 L^2}{24} \left( 2+\frac{3}{p^2 \uppi^2} \right) 
            \left[ 1+ \frac{2p^2\uppi^2 -6}{p^2\uppi^2 (2p^2\uppi^2+3)} \lambda \right],&  p = n 
        \end{cases},
    \end{aligned}
    \label{Eq:adx_quadG_baplhabeta}
\end{equation}
where $\quad \alpha = [0,0,2p], \beta = [0,0,2n], n,p>0$.
The result above is Taylor expanded near $\lambda=0$. According to Eq.~\eqref{Eq:s0_pertb}, we have
\begin{equation}
    \begin{aligned}
        &s_0^{\rm (3a)} 
        \approx - \rmi \frac{\gXe^3}{D^2} L^4  \times  \\
        &\sum_{n=1}^{\infty}{\sum_{p=1}^{\infty}{
            \frac{ b_{0,[0,0,2p]}b_{[0,0,2p],[0,0,2n]}b_{0,[0,0,2n]} }{
                \left[ \left( 2p\uppi + \frac{2\lambda}{2p\uppi}  \right)^2 -2\lambda \right]
                \left[ \left( 2n\uppi + \frac{2\lambda}{2n\uppi}  \right)^2 -2\lambda \right]
            }
        }}.
    \end{aligned}
    \label{Eq:adx_quadG_s03a_eq1}
\end{equation}
Taylor expand the above formula at $\lambda=0$ (up to first order), we get 
\begin{equation}
    \begin{aligned}
        s_0^{\rm (3a)} \approx& -\rmi \frac{\gXe^3 G_2^3 L^{10}}{16D^2 \uppi^{10}} 
        \left[ \left( 1-\frac{\lambda}{3} \right) \mathcal{S}_1 + \frac{\lambda}{2\uppi^2} \mathcal{S}_2 \right] \\
        &- \rmi \frac{19 \gXe^3 G_2^3 L^{10}}{59875200D^2} \left( 1-\frac{659}{5460}\lambda \right)
        + O(\lambda^2),
    \end{aligned}
    \label{Eq:adx_quadG_s03a_eq2}
\end{equation}
where
\begin{align}
    \mathcal{S}_1 &\equiv \sum_{n=1}^{+\infty}{\sum_{p=1}^{n-1}{ \frac{n^2 + p^2}{n^4 p^4 (n^2-p^2)^2} }} \approx 0.0375373, \\
    \mathcal{S}_2 &\equiv \sum_{n=1}^{+\infty}{\sum_{p=1}^{n-1}{ \frac{n^4 + 8 n^2 p^2 + p^4}{n^6 p^6 (n^2-p^2)^2} }} \approx 0.0892948. 
\end{align}
Summing $s_0^{\rm (3a)}$ and $s_0^{\rm (3b)}$ together, we finally get the third order correction:
\begin{equation}
    s_0^{(3)} \approx - \rmi \frac{\gamma_{\rm Xe}^3 G_2^3 L^{10} }{D^2} \chi_1 \left(  1 + \chi_2 \lambda \right) + O(\lambda^2), 
    \label{Eq:adx_quadG_s03} 
\end{equation}
with
\begin{align}
  \chi_1 &\equiv \frac{\mathcal{S}_1}{16 \uppi^{10}} + \frac{1}{23950080} \approx 6.68056 \times 10^{-8}, \\
  \chi_2 &\equiv \left(  \frac{15871}{326 918 592 000} - \frac{\mathcal{S}_1}{48 \uppi^{10}} + \frac{\mathcal{S}_2}{32 \uppi^{12}}  \right) / \chi_1 \notag \\
         &\approx 0.646886 . 
\end{align}

\subsection*{Systematic Error of NMR Gyroscope}
\label{sec:adx_NMRG_sysError}
In NMR gyroscope experiments, one often simultaneously measure the Larmor precession frequency of both \XeName{129} and \XeName{131} nuclear spin. 
The Larmor frequencies of these two Xe isotopes are
\begin{equation}
  \omega_{u} = - \gamma_{u} (B_0 + \delta B_{u}) - \Omega_{\rm rot},  
\end{equation}
where $u = 129~{\rm or}~131$, and $\delta B_{u} \equiv \Im{[s_0]}/\gamma_{u} - B_0$ is the frequency shift caused by nonuniform field. 
$\Omega_{\rm rot}$ is the laboratory reference frame's rotation angular velocity along $\hat{z}$ direction. 

One usually estimates the rotation speed by the following estimator~\cite{Zhang2023}:
\begin{equation}
  \begin{aligned}
  \Omega_{\rm rot}^{(2\omega)} \equiv& 
  \frac{\left| R_0 \omega_{131} \right| - \left|\omega_{129} \right|}
  {1 + \left| R_0 \right|} \\
  =& {\rm sgn}{[B_0]} \left( \Omega_{\rm rot}  + \frac{\gamma_{129}}{1 + |R_0|} b_{\rm A} \right),
  \label{Eq:adx_Omega_rot2_estimate}
  \end{aligned}
\end{equation}
where $b_{\rm A} \equiv \delta B_{129} - \delta B_{131}$ is called the differential field, and $R_0 \equiv \gamma_{129} / \gamma_{131} \approx -3.373417$. 
Obviously, this estimator introduces a systematic error proportional to $b_{\rm A}$. 
Utilizing the frequency shift formulas derived in the above section, it is straight forward to get the Eqs.~\eqref{Eq:deltaOmega_linG_1st}, \eqref{Eq:deltaOmega_quadG_1st} and \eqref{Eq:deltaOmega_quadG_3rd} (assuming $B_0 > 0$). 

\subsection*{Experiment Measurement of Wall Relaxation Rate}
\label{sec:adx_wallRelax_measurement}
According to literature~\cite{Volk1980a, Butscher1994}, the transverse relaxation rate $1/T_2$ of Xe spins mainly consists of two parts:
\begin{equation}
  \frac{1}{T_2} = \underbrace{c_1 n_{\rm Rb}(T)}_{\Gamma_{\rm collision}} 
                + \underbrace{c_2 \exp{\left( \frac{\bar{E}}{k_B T} \right)}}_{\Gamma_{\rm w}}, 
  \label{Eq:adx_Gamma2_fitModel}
\end{equation}
where $T$ is the cell temperature in degrees kelvin, $n_{\rm Rb}(T)$ is the Rb atom number density, $\bar{E}$ is a characteristic energy, $k_B$ is the Boltzmann's constant, and $c_1, c_2$ are constant coefficients. 
$\Gamma_{\rm collision}$ arises from the spin exchange collisions between Xe and Rb atoms and is proportional to Rb density. 
$\Gamma_{\rm w}$ comes from the wall interaction, which depends on cell temperature via an Arrhenius factor. 

The solid lines in Fig.~\ref{Fig:boundaryCond_measurement} is fitted curves using model Eq.~\eqref{Eq:adx_Gamma2_fitModel}, with $c_1, c_2, \bar{E}$ the fitting parameters. Rb density $n_{\rm Rb}(T)$ is calculated using the Rb vapor-pressure formula Eq.~(1) in Steck's handbook~\cite{SteckRb85_2021} (also see Alcock {\it et al.}~\cite{Alcock1984}) together with the Ideal Gas Law. The fitting result is shown in Tab.~\ref{Tab:adx_boundary_fitResult}. Quadrupole splitting of \XeName{131} is not observed in the experiment of Fig.~\ref{Fig:boundaryCond_measurement}. 

\begin{table*}[t]
  \renewcommand{\arraystretch}{1.5}
  \renewcommand{\tabcolsep}{8pt}
  \centering
  \captionsetup{width=1\textwidth, justification=centering}
  \caption{\textbf{Fitting result of the boundary conditions.} 
  The range indicated here corresponds to the 95\% confidence interval returned by the fitting algorithm.}
  \label{Tab:adx_boundary_fitResult}
  \begin{tabular}{c||c|c|c||c}
    \toprule
        &  $c_1~({\rm 10^{-15}~cm^3 \cdot s^{-1}})$  
        &  $c_2~({\rm 10^{-3}~s^{-1}})$  
        &  $\bar{E}~({\rm meV})$  
        &  $\Gamma_{\rm w}~({\rm s^{-1}}) @ 110\degC$ \\
    \colrule
    \XeName{129}  & $1.343 \pm 0.021$ & $1.23  \pm 0.34$ & $95.4 \pm 8.6$ & $0.0222 \pm 0.0084$\\
    \XeName{131}  & $0.382 \pm 0.028$ & $0.363 \pm 0.075$ & $165.6 \pm 6.4$ & $0.055 \pm 0.015$ \\
    \botrule
  \end{tabular}
\end{table*}

From Eq.~\eqref{Eq:s0_pertb}, the relaxation rate contributed from boundary condition is $\Gamma_{\rm w} = D \kappa_0^2$. Thus, for cubic cell, using the solution in section "\nameref{sec:adx_unperturbed_eigenmodes}" in the methods, we can estimate the value of $\lambda$ by $\GXe{w}$:  
\begin{equation}
  \Gamma_{\rm w} = \frac{6 \lambda D}{L^2}. 
  \label{Eq:adx_GammaWall_lambda_relation}
\end{equation}
The diffusion constant can be measured using the method described in Sec.~II~D of the Supplemental Material of Zhang {\it et al.}'s work~\cite{Zhang2023}. The result is $D_{129}=D_{131}=\left. (0.45 \pm 0.03)~{\rm cm^2 \cdot s^{-1}} \right.$. The inner side length of cell is $L = \left. (0.80 \pm 0.01)~{\rm cm} \right.$. Using these parameters, we get the boundary condition at $110\degC$ to be $\lambda_{129} = \left. (5.3 \pm 2.0) \times 10^{-3} \right.$ and $\lambda_{131} = \left. (13.0 \pm 3.8) \times 10^{-3} \right.$.

\section*{Data Availability}
The data that support the findings of this study are available from the corresponding author upon reasonable request. 

\section*{Author Contributions}
N.Z. designed and supervised the project. 
D.X. conducted preliminary research.
X.Z., J.H., and N.Z. performed the analytical derivation of key results.
X.Z. refined and analyzed the results.
X.Z. and N.Z. wrote the manuscript.
All authors contributed to reviewing the manuscript. 

\section*{Competing Interests}
The authors declare no competing interests.

\section*{Acknowledgments}
  We thank Prof. Dong Sheng for providing advice on the overall structure of this article and assisting in the production of the vapor cell used in Fig.~\ref{Fig:boundaryCond_measurement}. We also thank Dr. Sheng Zhang for his suggestions for improving the presentation of this article. 
  This work is supported by the NSFC (Grants No. U2030209, No. 12088101, and No. U2230402) and the China Postdoctoral Science Foundation (Grants No. GZB20240462).

\section*{References}
\bibliographystyle{naturemag}
\bibliography{FreqShift_TorreyEq.bib}

\end{document}


\title{Supplemental Material for \\ Frequency Shift Caused by Nonuniform Field and Boundary Relaxation in Magnetic Resonance and Comagnetometers}
\author{Xiangdong Zhang}
\affiliation{Beijing Computational Science Research Center, Beijing 100193, People's Republic of China}%
\affiliation{College of Physics and Optoelectronic Engineering, Shenzhen University, Shenzhen 518060, China}%
\author{Jinbo Hu}
\affiliation{Beijing Computational Science Research Center, Beijing 100193, People's Republic of China}%
\author{Da-Wu Xiao}
\affiliation{Beijing Computational Science Research Center, Beijing 100193, People's Republic of China}%
\author{Nan Zhao}
\email{nzhao@csrc.ac.cn}
\affiliation{Beijing Computational Science Research Center, Beijing 100193, People's Republic of China}%
\date{\today}

\maketitle

\tableofcontents

\renewcommand\thefigure{S\arabic{figure}} 
\renewcommand\theequation{S\arabic{equation}} 
\renewcommand\thetable{S\arabic{table}} 
\renewcommand\thesection{Supplementary Note \arabic{section}}
\setcounter{figure}{0}
\setcounter{equation}{0}
\setcounter{table}{0}

\vspace{2em}

(Note: In order not to introduce confusion, in this Supplemental Material, as well as maintext, subscript $m, n, p$ will always represent a single integer index, while a Greek subscript such as $\alpha, \beta, \gamma$ represents multiple indices.)

\section{About the Odd-Parity of Magnetic Field\label{sec:adx_odd_parity}}
The frequency shift caused by nonuniform magnetic field can be estimated by the $\rm 1^{st}$ and $\rm 3^{rd}$ order terms of Eq.~\eqref{Eq:s0_pertb} in maintext. 
However, when the solution domain $V$ is symmetric and the nonuniform part of $B_z$ has odd-parity along some directions, the integration $b_{00}$ and $b_{0\alpha} b_{\alpha \beta} b_{\beta 0}$ will vanish, leading to a zero frequency shift. In this case, the frequency shift effect of the nonuniform part of $B_z$ is suppressed by symmetry, we need to consider the effect of transverse components $B_x$ and $B_y$, which is smaller but not negligible now. To count the effect of transverse components, one needs to replace the $b_{\alpha \beta}$ in Eq.~\eqref{Eq:s0_pertb} of maintext by the $b^{(\rm tot)}_{\alpha\beta}$ in Eq.~\eqref{Eq:adx_balphabeta_tot_formula}. 

As an example of symmetry suppression, consider a cubic domain $V$, where the eigenmodes $\phi_{\alpha}(\bfr)$ in the definition of $b_{\alpha \beta}$ (Eq.~\eqref{Eq:adx_def_bMatrix2}) have a factorized form as stated in Eq.~\eqref{Eq:adx_eigenMode_sepForm}. From Eqs.~\eqref{Eq:adx_phiz_ansatz} and \eqref{Eq:adx_delta_solution}, we can see the single axis eigenmodes have either odd or even parity:
\begin{equation}
  \begin{cases}
    \phi_p(z) = + \phi_p(- z),&  p=0,2,4,\cdots \\
    \phi_p(z) = - \phi_p(- z),&  p=1,3,5,\cdots
  \end{cases}
\end{equation}
Thus, for a $B_1(\bfr)$ with odd-parity along $z$ axis, which satisfies $B_1(z) = - B_1(-z)$, we have
\begin{equation}
  \begin{aligned}
    b_{\alpha \beta} =
    b_{mnp,m'n'p'} =& \int_{-L/2}^{+L/2}{\rmd x \phi_m(x) \phi_{m'}(x)} 
                  \int_{-L/2}^{+L/2}{\rmd y \phi_n(y) \phi_{n'}(y)}
                  \int_{-L/2}^{+L/2}{\rmd z \phi_p(z) \phi_{p'}(z) B_1(z)}  \\
              =& \delta_{m m'} \delta_{n n'} 
              \int_{-L/2}^{+L/2}{\rmd z \phi_p(z) \phi_{p'}(z) B_1(z)}
  \end{aligned}.
\end{equation}
$b_{\alpha \beta}$ is nonzero only when $m=m', n=n'$, and $p, p'$ have different parity. So, $b_{00}=0$ and $b_{0 \alpha}b_{\alpha \beta}b_{\beta 0} = 0, \forall \alpha, \beta$. For $B_1(\bfr)$ satisfying $B_1(x) = - B_1(-x)$ or $B_1(y) = - B_1(-y)$, we have the same conclusion.

\section{Compare Linear Gradient Frequency Shift with Result in Literature\label{sec:adx_cmp_literature}}

As mentioned in maintext, the Eq.~(32) of \citeZheng is twice as large as our result Eq.~\eqref{Eq:freqShift_linG_1st} of maintext. In this section, we will show that there is a typo in \citeZheng, and the derivation in this work is actually consistent with the theory in \citeZheng. 

Equation~(32) of \citeZheng is:
\begin{equation}
\delta\omega= \frac{\gamma^2 L^2}{12 \omega_0}\left(
  \left| \vec{\nabla} B_x \right|^2 + \left| \vec{\nabla} B_y \right|^2
\right).
\label{Eq:adx_Zheng2011_Eq32}
\end{equation}
As stated in \citeZheng, this result applies to cubic domain with non-relaxation boundary condition ($\lambda=0$). The $L$ here has the same meaning as in our work. $\gamma$ corresponds to $\gXe$, and $\omega_0$ corresponds to $\gXe B_0$. Using the linear gradient field expression Eq.~\eqref{Eq:linear_gradient_3Ddistribution} in maintext, we have 
$|\vec{\nabla} B_x| = |\vec{\nabla} B_y| = G_1/2$. Thus, Eq.~\eqref{Eq:adx_Zheng2011_Eq32} becomes
\begin{equation}
  \delta\omega= \frac{\gXe G_1^2 L^2}{24 B_0},
\end{equation}
which is twice as large as Eq.~\eqref{Eq:freqShift_linG_1st} of maintext.

However, as stated in \citeZheng, Eq.~\eqref{Eq:adx_Zheng2011_Eq32} is derived by substituting Eq.~(24) of \citeZheng into Eq.~(30) of \citeZheng. The Eq.~(24) of \citeZheng is
\begin{equation}
  J_x(\omega) = \frac{8 L^2}{\uppi^4} \sum_{n=1,3}^{\infty}{
    \frac{1}{n^4} \frac{1}{\frac{n^2 \uppi^2 D}{L^2}+ i \omega}
  }.
\end{equation}
In the fast diffusion limit with high pressure (as stated in \citeZheng, $\omega L^2 / 32 \uppi D \gg 1$), the imaginary part of $J_x(\omega)$ is simplified to
\begin{equation}
    \Im{\left[ J_x(\omega) \right]} = - \frac{8L^2}{\uppi^4 \omega}
    \sum_{n=1,3}^{\infty}{\frac{1}{n^4} \frac{1}{n^4 \xi^2 + 1}}
    \approx  - \frac{L^2}{12 \omega} + O\left(\xi^{3/2}\right),
  \label{Eq:adx_Zheng2011_ImJx1}
\end{equation}
where $\xi^2 \equiv \left(\uppi^2 D / \omega L^2 \right)^2 \ll 1$, and the approximation above represents a Taylor expansion at $\xi=0$. Substitute Eq.~\eqref{Eq:adx_Zheng2011_ImJx1} into Eq.~(30) of \citeZheng, we have
\begin{equation}
  \delta \omega = - \frac{\gamma^2}{2} \left( 
    \left| \vec{\nabla} B_x \right|^2 + \left| \vec{\nabla} B_y \right|^2
    \right) \Im{[J_x(\omega)]}
    \approx \frac{\gamma^2 L^2}{24 \omega}\left( 
      \left| \vec{\nabla} B_x \right|^2 + \left| \vec{\nabla} B_y \right|^2
      \right),
\end{equation}
which is half as large as Eq.~\eqref{Eq:adx_Zheng2011_Eq32}, and is consistent with our result Eq.~\eqref{Eq:freqShift_linG_1st} of maintext.

We can also check the Eq.~(33) of \citeZheng, by first re-expand Eq.~\eqref{Eq:adx_Zheng2011_ImJx1} in the fast diffusion limit with low pressures (expand at $\xi \rightarrow +\infty$):
\begin{equation}
  \Im{\left[ J_x(\omega) \right]} = - \frac{8L^2}{\uppi^4 \omega}
  \sum_{n=1,3}^{\infty}{\frac{1}{n^4} \frac{1}{n^4 \xi^2 + 1}}
  \approx  - \frac{17 \uppi^4 L^2}{20160 \omega \xi^2} + O\left(\frac{1}{\xi^4}\right).
\label{Eq:adx_Zheng2011_ImJx2}
\end{equation}
Substitute into Eq.~(30) of \citeZheng, we get
\begin{equation}
  \delta \omega = - \frac{\gamma^2}{2} \left( 
    \left| \vec{\nabla} B_x \right|^2 + \left| \vec{\nabla} B_y \right|^2
    \right) \Im{[J_x(\omega)]}
    \approx \frac{17 \omega \gamma^2 L^6}{40320 D^2}\left( 
      \left| \vec{\nabla} B_x \right|^2 + \left| \vec{\nabla} B_y \right|^2
      \right),
\end{equation}
which is also half as large as the Eq.~(33) of \citeZheng, and has a different exponent of $L$ (Eq.~(33) is proportional to $L^8$). By dimensional analysis, we can confirm that the $L^6$ dependency is correct.

\bibliographystyle{apsrev4-2}
\bibliography{FreqShift_TorreyEq.bib}